\documentclass[fleqn,12pt]{article}
\usepackage{epsfig}

\usepackage{graphicx}
\usepackage{amssymb}
\usepackage{makeidx}
\usepackage{graphicx}
\usepackage{multicol}

\newcommand{\nn}{\nonumber\\ }
\newcommand{\beq}{\begin{eqnarray}}
\newcommand{\eeq}{\end{eqnarray}}
\newcommand{\be}{\begin{eqnarray}}
\newcommand{\ee}{\end{eqnarray}}
\newcommand{\BQ}{\begin{equation}}
\newcommand{\EQ}{\end{equation}}
\newcommand{\BQA}{\begin{eqnarray}}
\newcommand{\EQA}{\end{eqnarray}}
\newcommand{\NN}{\nonumber \\}

\newcommand{\z}{z_\perp}

\newcommand{\rr}{r_\perp}

\def\labe{\label}
\def\simge{\mathrel{%
   \rlap{\raise 0.511ex \hbox{$>$}}{\lower 0.511ex \hbox{$\sim$}}}}
\def\simle{\mathrel{
   \rlap{\raise 0.511ex \hbox{$<$}}{\lower 0.511ex \hbox{$\sim$}}}}
\def\bigs{\mathrel{
   \rlap{\raise 0.531ex \hbox{$>$}}{\lower 0.531ex \hbox{$<$}}}}

\def\grad{\nabla}                               % gradient
\def\del{\partial}                              % synonym for \partial

\newcommand{\kk}{k_\perp}

\begin{document}

\begin{titlepage}

\begin{flushright}
SACLAY-T02/181 
\end{flushright}

\vspace{1cm}

\begin{center}
{\LARGE\sf A Gaussian effective theory for gluon saturation}

\vspace{0.9cm}

{\large Edmond Iancu$^{\rm a}$, Kazunori Itakura$^{\rm a,b}$, 
and  Larry McLerran$^{\rm c}$}\\

\vspace{5mm}

{\it $^{\rm a}$~Service de Physique Theorique, CE Saclay, F-91191 
        Gif-sur-Yvette}

\vspace{0.1cm}

{\it $^{\rm b}$~RIKEN BNL Research Center, BNL, Upton NY 11973}

\vspace{0.1cm}

{\it $^{\rm c}$~Nuclear Theory Group, Brookhaven National Laboratory,
        Upton, NY 11973  } 

\vspace{0.5cm}

\end{center}

\vspace{0.5cm}

\begin{abstract}
We construct a Gaussian approximation to the effective theory for
the Colour Glass Condensate which describes correctly the gluon
distribution both in the low density regime at high transverse momenta
(above the saturation scale $Q_s$),
and in the high density regime below $Q_s$, and provides a simple interpolation
between these two regimes. At high momenta, the effective theory reproduces 
the BFKL dynamics, while at low momenta, it exhibits gluon
saturation and, related to it, colour neutrality over 
the short distance scale $1/Q_s \ll 1/\Lambda_{QCD}$.
Gauge--invariant quantities computed within this approximation
are automatically infrared finite.

\end{abstract}

\end{titlepage}
\section{Introduction}
\setcounter{equation}{0}

There has been significant progress towards the understanding of the
high energy, or small--$x$, limit of QCD in the recent years, and part of
this progress is associated with the construction of an effective theory for
the small--$x$ gluons in the  hadron light--cone wavefunction:
the  effective theory for the {\it Colour Glass Condensate} (CGC)
\cite{MV,K96,JKMW97,KM98,LM00,JKLW97,PI,SAT,FB} (see also Refs. 
\cite{Cargese,AMlectures} for recent reviews and more references).
This is an {\it effective} theory since it applies to gluons with a given
value of the longitudinal momentum fraction $x$, with $x\ll 1$,
and is obtained after integrating out the gluons with larger values
of $x$ within QCD perturbation theory. 

The general strategy behind %underlying
this construction is reminiscent of the Born--Oppenheimer approximation
(i.e., there is a separation of scales
 between ``fast'' and ``slow'' degrees of freedom),
and the resulting mathematical description is that of a {\it glass}.
The ``fast'' degrees of freedom are the gluons (or, generally, partons)
with larger values of $x$, which do not participate directly in the
high--energy scattering (because of their large relative rapidities),
but act as sources for the small--$x$ gluons which produce the 
scattering. The ``slow'' degrees of freedom are the small--$x$ gluons
themselves, which have low rapidities (of the same order as the external
projectile), but large densities in the impact parameter space (the transverse
plane), since their radiative production is enhanced at small--$x$.

This whole partonic system is characterized by weak coupling %``constant''
--- since the high gluon density acts as a hard scale for the running
of the QCD coupling $\alpha_s$ ---, but strong interactions with the
small--$x$ gluons, because these latter have densities which are parametrically
of order $1/\alpha_s$. The crucial simplification, which allows for 
the all--order resummation of the density--enhanced interactions,
is that the small--$x$ gluons can be treated
in the classical approximation \cite{MV}, because of their 
large occupation numbers. That is, the small--$x$ gluons can be
described as the classical colour fields radiated 
by fast moving ``colour sources''
which represent the partons with larger values of $x$.
Then, the all--order resummations alluded to before amount
to treating the non--linear effects associated with
these classical fields {\it exactly\,}, both in
the classical field equations \cite{MV,K96,JKMW97,KM98}, 
and in the quantum calculations involved in the
construction of the effective theory \cite{JKLW97,PI}.

%Then, the strong interactions alluded to before appear as non-linear effects
%in the classical field equations, and also as ``background field'' effects
%in the quantum evolution of the classical effective theory.
% calculations which permit the construction of the 
%effective colour source at small--$x$.

A second simplification, coming from the separation of scales between
``fast'' and ``slow'', is that the internal 
dynamics of the fast partons can be neglected over the
characteristic time scales for the dynamics at small $x$.
This is the premise of the ``glassy'' description : 
For the purposes of the small--$x$ physics, the fast partons can be 
treated as a {\it frozen} configuration of classical colour sources,
which is {\it random}, in the sense that it can be any of the
configurations allowed by the dynamics of the fast partons, and which
must be  {\it averaged over} in the calculation of physical quantities.
This averaging is necessary also
to restore gauge symmetry: on the average, the colour charge
density must vanish at any point. This means that the {\it weight function}
for this averaging, i.e., the probability density for having a given
configuration of the colour sources,
 must be a gauge--invariant functional of the colour charge density.
Further characterizing this  functional, and deriving an explicit
expression for it, will be our general objective in this paper.

This weight function encodes all the information about the colour source
and its correlations, as inherited from the quantum dynamics
at larger values of $x$. When $x$ is further decreased, the structure of the
classical field equations does not change: these are always the Yang--Mills
equations with a colour source in their right hand side. What do change,
are the correlations of the colour source, that is, its functional
weight function. This change can be computed in perturbation theory,
and expressed as a functional {\it renormalization group equation} (RGE)
for the weight function \cite{JKLW97,PI}. 
The initial condition for this equation 
is non-perturbative, and requires a model.
But once this  is specified, then the
effective theory at scale $x$ is completely
determined by the solution to the RGE.

Even though the general solution to this equation is not known 
(this is a complicated non--linear equation \cite{PI}; 
see, however, \cite{PATH}),
its physical implications are well understood, at least, in some
limiting kinematical regimes, where approximate solutions have been 
obtained \cite{SAT,Cargese}. One has thus identified two important physical
regimes:  (I) a low--density regime at high transverse momenta (for fixed $x$),
where the non--linear effects can be expanded out, and
the RGE reduces to the BFKL evolution \cite{BFKL}, and  (II)
a high--density regime at low transverse momenta, where the 
non--linearities are essential, and the RGE predicts
{\it gluon saturation} \cite{GLR,MQ85,BM87}. The transition between
 these two regimes occurs at  transverse momenta of the order of the
``saturation scale'' $Q_s(x)$, which is the typical momentum of the
saturated gluons, and grows like a power of $1/x$
\cite{AM2,SCALING,MT02,DT02}. %  (and Refs. therein).
%\cite{AM2,SCALING,MT02}

These conclusions are corroborated %obtained in this way 
by analytic \cite{K,LT99,SCALING,FB} and numerical \cite{LT99,AB01,Motyka,LL01} 
studies of the Balitsky--Kovchegov (BK) 
equation \cite{B,K}, which describes deep inelastic scattering at high energy, 
and is encoded too in the RGE  \cite{PI}.
Originally, the BK equation has been derived in a
different approach \cite{AM0,AM3,B,K,AM2,Braun,W,AM01}
(see also Refs. \cite{GLR,MQ85} for earlier versions of this equation),
which focuses on the %follows the evolution of the
observables through which an external projectile 
couples to a hadronic target, rather than on
the wavefunction of the target itself.

Our purpose in this paper is to use the conclusions of these previous
studies to construct an approximate solution to the RGE
which is simply a {\it Gaussian} in the colour charge density $\rho$.
Clearly, having an explicit expression for the weight function,
which is moreover a Gaussian, will greatly facilitate the use
of the  CGC effective theory in practical calculations.
In fact, with such a weight function, most calculations
within the effective theory can be done quasi--analytically\footnote{Note
that, even with a Gaussian weight function, the CGC effective theory
remains non--trivial, since the classical field equations
(i.e., the Yang--Mills equations with colour source $\rho$) are non--linear.
Still, as we shall remind in Sect. 2, the exact solution to these
equations is known in analytic form.}.
Besides, a Gaussian weight function can be easily implemented 
in numerical simulations, like those performed in Ref. \cite{KV},
where the effective theory is used to study nucleus--nucleus
collisions on a lattice.

The adequacy of a Gaussian approximation may look at first surprising, given
the highly non--linear structure of the RGE.
%, and the fact that our main
%interest in using this equation is to
%describe the high density regime in which the non--linearities are 
%truly important.
By definition, a Gaussian approximation retains only the 2--point  
correlation function $\langle\rho\rho\rangle$ among all the $n$--point
functions of the colour sources. 
Thus, this is a reasonable approximation provided ({\it a\,})
one disposes of a good approximation for the 2--point function, and
({\it b\,}) the higher $n$--point correlations are relatively weak.
But in the RGE, the evolution of the 2--point function
is generally mixed with that of the higher $n$--point functions, so
it is not  a priori clear how to construct a sensible approximation 
for the 2--point function alone (and thus a Gaussian weight function).
 It is even less clear whether such an approximation 
can be relevant for the non--linear regime, where higher correlations
may be important as well.

Still, as we shall demonstrate below, a Gaussian weight
function can accommodate both the
BFKL evolution of the gluon distribution at high transverse momenta,
and the phenomenon of gluon saturation at low transverse momenta.
It is only the transition between these two regimes that cannot be
accurately described in this way. (But our approximation will provide
a smooth interpolation between these limiting regimes.) 
To understand when, and  to which extent, is a Gaussian approximation
expected to work, let us make a few remarks that will get
substantiated in the subsequent analysis:

In the low--density regime at
 high transverse momenta $\kk \gg Q_s(x)$,
the evolution generated by the RGE does not couple correlation functions
with different number of external legs. In particular, the 2--point function
$\langle\rho\rho\rangle$ satisfies a closed equation, which is moreover
linear \cite{JKLW97}. This is the BFKL equation \cite{BFKL}. By solving
this equation, one can construct a Gaussian weight function which incorporates
correctly the BFKL dynamics for both the unintegrated gluon distribution,
and deep inelastic scattering.

In the saturation regime at low momenta $\kk \ll Q_s(x)$,
the colour sources and the radiated fields are typically strong 
(their amplitudes are parametrically of order $1/g$), so the dynamics is
fully non--linear. Remarkably, however, %in this regime
the non--linear effects drop out completely from the RGE. 
This is related to the specific way how these effects enter the 
RGE in the first place \cite{PI} :
they enter only via {\it Wilson lines}, that is, path--ordered exponentials
of the classical fields. At saturation, where the fields  are strong, 
these complex exponentials oscillate  rapidly and average
to zero, together with all their correlations, when probed over transverse
distances of the order of the saturation length $1/Q_s(x)$, or larger.
Thus, in this high--density regime, the RGE becomes effectively 
{\it quadratic}\footnote{In the same sense, for instance, as the
Hamiltonian for a quantum--mechanical harmonic oscillator is
a quadratic operator.}, and can be trivially solved. The resulting
 weight function is then truly a Gaussian \cite{SAT}.
%As we shall see, at high momenta a Gaussian weight function

We see an interesting duality emerging at saturation: this strong
field regime allows for a description in terms of a Gaussian weight
function, so like a free theory.

Another important feature at saturation is {\it colour neutrality} 
\cite{SAT,FB,Cargese,AM02} :
The saturated gluons shield each other in such a way
that their global colour charge vanishes when integrated over
a disk of radius  $1/Q_s(x)$, or larger. This is expressed
by the following property of the charge--charge correlator:
$\langle\rho(\kk)\rho(-\kk)\rangle\propto \kk^2$ 
for momenta $\kk \ll Q_s(x)$. % \cite{SAT,FB,Cargese,AM02}.
This property has crucial implications:

First of all, it is  synonymous of saturation: As we shall see,
$\langle\rho(\kk)\rho(-\kk)\rangle/\kk^2$ is 
essentially the unintegrated gluon distribution.
Thus, the fact that the charge--charge correlator 
is proportional to $\kk^2$ at low momenta, means that
the unintegrated gluon distribution is independent of $\kk$ 
(up to logarithms), which is saturation indeed.
In fact, we shall see that
the scale $Q_s(x)$ is primarily generated (by the
non-linear effects in the quantum evolution) as the typical
scale for colour neutrality. Then this becomes also the
``saturation scale'' because of the connection between the
distribution of the colour sources and that of the small--$x$
gluons alluded to above.

Second, colour neutrality ensures that gauge--invariant
quantities come out infrared finite
when computed in the effective theory. % come out infrared finite.
This should be contrasted with the original  McLerran--Venugopalan (MV) 
model \cite{MV}, in which the colour sources are assumed to be uncorrelated,
so the corresponding charge--charge correlator 
$\langle\rho(\kk)\rho(-\kk)\rangle$ is independent of $\kk$.
Because of that, the calculation of physical quantities 
like the  gluon distribution function \cite{JKMW97},
 or the dipole--hadron scattering amplitude \cite{Cargese},
within this model meets with logarithmic infrared divergences.
It has been proposed by Lam and Mahlon \cite{LM00}
to remove these divergences by imposing 
colour neutrality on the size of the nucleon, that is, by effectively
taking into account the non--perturbative mechanism of confinement.
In Ref. \cite{KV}, this suggestion has been implemented
in numerical simulations, but the results turn out to be quite 
sensitive to the poorly known non--perturbative physics.
As we shall show in this paper, when the colour neutrality due
to saturation is taken into account,
all such infrared divergences are eliminated 
already at the perturbative scale $1/Q_s(x)$.

Other effects which can be attributed to
colour neutrality at saturation, and which illustrate the lack of infrared
sensitivity of the effective theory, are the suppression of the
``infrared diffusion'' in the quantum evolution towards small $x$ (this
has been verified in the  numerical simulations 
of the BK equation \cite{Motyka}), and the predominance of the
short-range scattering in the evolution of the total cross-sections
with increasing energy \cite{FB}.

Note that colour neutrality is a weaker property than confinement:
Although the global colour charge vanishes indeed, the multipolar
moments (starting with the dipolar one) are still  non--zero,
and generate long range interactions. But the latter
die away sufficiently fast to ensure the infrared finiteness
of all observables for high--energy scattering (at fixed impact
parameter). The information about confinement is needed
only in the calculation of the total cross--section,
since this requires an integration over {\it all\,} 
impact parameters \cite{FB,LR90,KW02}.
Still, as shown in Ref. \cite{FB} on the example of the dipole--hadron 
scattering, by  combining the non--linear
evolution equations with a minimal assumption about confinement in the
initial conditions, one can  compute the rate for the increase 
of the total cross--section with the energy, and conclude that
$\sigma_{\rm tot}$ saturates the Froissart bound \cite{Froissart}
at very high energy.
Similar conclusions have been reached in Refs. \cite{LR90,KL02}.

In view of the previous considerations, our general
strategy in this paper should be clear by now:
We shall construct our Gaussian approximation for the weight function
by requiring its kernel $\langle\rho\rho\rangle$ to satisfy the BFKL
equation at high momenta, to describe colour neutrality at low momenta,
and to interpolate smoothly between these two limiting regimes, 
with the change in behaviour occurring at the saturation scale.
The last condition requires a saturation criterion for the
charge--charge correlator, that we shall formulate, and prove to be
equivalent to the other criteria used in the literature. For practical
purposes, we shall propose also a simplified form of the kernel,
in which the general solution to the BFKL equation is replaced by
its ``geometric scaling'' approximation \cite{SCALING}, valid
up to momenta $\kk\sim Q_s^2(x)/\Lambda_{QCD}$.

To test the quality of our approximation, we shall  compute the
unintegrated gluon distribution and the dipole--hadron scattering 
amplitude, and find the expected % limiting
results at both high and low momenta.
In particular, from the kernel of the Gaussian, we shall deduce a simple
analytic formula for the unintegrated gluon distribution,
which has a {\it scaling} form, i.e., it is a function of $\kk/Q_s(x)$,
and interpolates smoothly between saturation at $\kk < Q_s(x)$ and
BFKL behaviour at $\kk > Q_s(x)$. 

Also, we shall derive
a phenomenologically useful formula for the dipole--hadron scattering
amplitude, which properly incorporates the constraints of 
geometric scaling \cite{geometric} 
(both in the saturation regime, and in the scaling window \cite{SCALING}
above $Q_s (x)$), and the colour neutrality due to saturation.
This  analytic formula can be easily applied to the description of 
electron--hadron deep inelastic scattering, as an alternative to
the more phenomenological parametrizations proposed
in Refs. \cite{GBW99,BGBK}.

At this point, let us note an important difference in the mechanism of
saturation between the Gaussian effective theory that we shall construct
and the original MV model \cite{MV}, which has a Gaussian weight function too: 
In the MV model, the colour sources are uncorrelated, and 
the saturation arises exclusively via the non--linear
effects in the Yang--Mills dynamics of the classical fields. 
%This is {\it classical} saturation. 
By contrast, in the effective theory
which includes quantum evolution, the saturation is the
consequence of the long--range spatial correlations between the colour
sources, and, as such, it is encoded already in the weight function.
This explains why, in our case, the  unintegrated gluon distribution
can be simply proportional to the  kernel of the Gaussian even in the
non--linear regime at saturation.

This paper is organized as follows:
In Section 2, we shall succinctly review the effective theory
for the CGC, and introduce the notations and conventions to be used throughout.
In Section 3, we show that Gaussian approximations for the weight function
can be formally generated through {\it mean field  approximations}
to the RGE. In Section 4, we briefly describe the MV model 
\cite{MV,K96,JKMW97,LM00}, %which has a Gaussian weight function too, and 
which will serve as a basis of comparison for the following developments.
In Section 5, we discuss the simplifications which occur in the RGE
in the limiting situations where the transverse momenta are either very large,
or very small, compared to the saturation scale. This justifies
the use of a  Gaussian approximation in these regimes, at least.
In Section 6,
we construct an approximation for the charge--charge correlator
which interpolates smoothly between the limiting behaviours found previously.
This specifies the Gaussian effective theory completely.
In Section 7, we use this effective theory to first derive a simple 
expression for the unintegrated gluon distribution, and then
compute the dipole--hadron scattering amplitude.
Our conclusions are %presented
summarized in  Section 8.

\section{The Colour Glass Condensate}
\setcounter{equation}{0}

\subsection{The classical effective theory}

Let us summarize here the equations at the basis of the effective theory
for the CGC. The  classical field obeys the Yang--Mills
equations with a colour current due to the fast 
partons\footnote{We recall that, in light--cone coordinates, and with our
present conventions, $x^+\equiv (t+z)/\sqrt{2}$ plays
 the role of the light--cone 
time, while $x^-\equiv (t-z)/\sqrt{2}$ is the longitudinal coordinate.
Correspondingly, $k^-$ is the light--cone energy, and $k^+$
is the longitudinal momentum; thus, $x=k^+/P^+$, where  $k^+$
refers to a parton in the hadron, while 
$P^+$ denotes the total longitudinal momentum of the hadron.} :
\beq
(D_{\nu} F^{\nu \mu})_a(x)\, =\, \delta^{\mu +} \rho_a(x^-,x_\perp)\,.
\label{cleq}
\eeq

The structure of the colour current in the r.h.s. 
is fixed by the kinematics: {\it i\,}) The current has just a plus component
($J^\mu_a\simeq \delta^{\mu +}\rho_a$) since the fast partons
move nearly at the speed of light in the positive $z$ (or positive $x^+$) 
direction. {\it ii\,}) By Lorentz contraction, the support of the colour
charge density $\rho_a$ is concentrated near the light--cone, i.e.,
near $x^-=0$. % (see below for details on the longitudinal structure of $\rho$). 
{\it iii\,}) By Lorentz time dilation, $\rho_a$ is independent of the
light--cone time $x^+$, i.e., it is ``frozen'', according to the terminology 
used in the Introduction. 

All the approximations involved in writing  eq.~(\ref{cleq}) are
consistent with the quantum evolution towards small--$x$ in 
the leading--log
approximation, since this evolution preserves the separation 
of scales in longitudinal momenta (and therefore also in time).

The observables that can be computed in the effective theory
are the gauge--invariant correlations of the gauge fields.
These are first evaluated on the solution  $A^\mu[\rho]$ 
to eq.~(\ref{cleq}), and the
result is then averaged over all the configurations
of $\rho$, with a gauge--invariant weight function $W_\tau[\rho]$
which depends upon $x$ via the rapidity variable
 $\tau\equiv \ln(1/x)$.

For instance, we shall need later the scattering amplitude for the
 high energy scattering between a ``colour dipole'' (say, a
quark--antiquark pair in a colourless state) and the hadron.
This quantity is a physical observable (at least, indirectly)
in the sense that it enters the formula for the $F_2$ structure
function at small $x$ \cite{AM0,NZ91}.
In the eikonal approximation, the scattering amplitude
is obtained as \cite{BH,B} :
\beq\label{Ntau}
{\cal N}_\tau(x_{\perp},y_{\perp})\,=\,1\,-\,
S_\tau(x_{\perp},y_{\perp}),\qquad S_\tau(x_{\perp},y_{\perp})
\,\equiv\,\frac{1}{N_c}\,
\Big\langle {\rm tr}\Big(V^\dagger(x_{\perp}) V(y_{\perp})\Big)
\Big\rangle_\tau,\nonumber\\
\Big\langle {\rm tr}\Big(V^\dagger(x_{\perp}) V(y_{\perp})\Big)
\Big\rangle_\tau\,=\, \int { D}\rho\, \,W_\tau[\rho]\,\,{\rm tr}\Big
(V^\dagger_{x_{\perp}}[\rho] V_{y_{\perp}}[\rho]\Big),
\qquad\qquad\qquad\qquad\eeq
where $V^\dagger(x_{\perp})$ and $V(y_{\perp})$ are Wilson lines
describing the interactions between the fast moving quark, or
antiquark, from the dipole and the colour field in the hadron:
\be\labe{v}
V^\dagger(x_{\perp})\,\equiv\,{\rm P} \exp
 \left \{ig \int dx^-\,A^+_a (x^-,x_{\perp})t^a
 \right \},\ee
and $A^+_a\equiv A^+_a[\rho]$ is the projection of
the classical colour field along the trajectory of the quark
(or the antiquark). The Wilson line correlator in eq.~(\ref{Ntau})
is written in a suitable gauge in which the component $A^+$ is non--zero,
but the result is actually gauge--independent, since this correlator
can be completed to a Wilson loop by joining the points $x_{\perp}$ 
and $y_{\perp}$ via Wilson lines in the transverse planes at $x^-=\pm\infty$ 
(where the
transverse fields $A^i$ can be chosen to  vanish
in any gauge with $A^+\ne 0$).

Explicit calculations are most conveniently performed in
 the {covariant gauge} $\del_\mu A^\mu_a =0$, since in this gauge the 
solution $A^\mu$ to the classical equation (\ref{cleq})
has only one non-trivial component: $A^+_a=\delta^{\mu +}\alpha_a$,
with $\alpha_a$ satisfying the two--dimensional Poisson equation:
\beq\label{Poisson}
- \nabla^2_\perp \,\alpha_a(x^-,x_{\perp})\,=\,\rho_a(x^-,x_{\perp}).
\eeq
This simplification reflects the fact that the classical field
radiated by $\rho$ has only one non--trivial field strength,
namely the ``electric field'' $F^{+i}_a$,
which in the covariant gauge is computed as $-\del^i\alpha_a$.
The corresponding result in other gauges is then obtained
via the appropriate gauge rotation. Particularly interesting for what
follows is the result in the light--cone (LC) gauge $A^+_a=0$, which
reads (in matrix notations; see \cite{PI,Cargese} for details) :
\be\label{Fi+}
{F}^{+i}(x^-,x_{\perp}) %\,\equiv\, \partial^+
%{\cal A}^{i}(x^-,x_{\perp})
\,=\,U\,\big(-\del^i\alpha\big)\,U^\dagger\,,\ee
where 
\be
U^{\dagger}(x^-,x_{\perp})\,=\,
 {\rm P} \exp
 \left \{ig \int_{-\infty}^{x^-} dz^-\,\alpha (z^-,x_{\perp})
 \right \},\labe{UTA}
\ee
is the matrix for the gauge rotation from the covariant gauge to the
LC gauge. This is essentially the same Wilson line as for
the quark scattering in the eikonal approximation, eq.~(\ref{v}).
Note that eqs.~(\ref{Poisson})--(\ref{UTA}) provide an explicit
expression for the electric field
in the LC gauge in terms of the colour source $\rho_a$
in the covariant gauge. In this sense, the solution to the classical
field equations  (\ref{cleq}) is known {\it exactly} (in any gauge).

Eq.~(\ref{Poisson}) can be easily inverted :
\be\label{alpha}
\alpha_a (x^-,{ x}_\perp)&=&\int d^2y_\perp\,
\langle x_\perp|\,\frac{1}{-\grad^2_\perp}\,|y_\perp\rangle\,
\rho_a  (x^-,{ y}_\perp)\nn&=&
\int \frac{d^2y_\perp}{4\pi}\,
\ln\frac{1}{({x}_\perp - {y}_\perp)^2\Lambda^2}\,\,
\rho_a  (x^-,{ y}_\perp).\ee
The infrared cutoff $\Lambda$ is necessary to invert the
Laplacian in two dimensions, but, as we shall see, it disappears
in the calculation of physical quantities.
From  eq.~(\ref{alpha}), it is clear that the ``Coulomb field'' 
$\alpha(x^-,{ x}_\perp)$ is time--independent and localized near
$x^-=0$, like the colour source.
%and so is also the ``electric field'' $F^{+i}$ in any gauge.

To summarize, to compute the dipole--hadron scattering amplitude (\ref{Ntau})
within the effective theory,
one has to replace $A^+_a\rightarrow \alpha_a$ in eq.~(\ref{v}),
with $\alpha_a$ given by eq.~(\ref{alpha}),
and then perform the functional integral over $\rho$ in the second line
of eq.~(\ref{Ntau}), with the weight function $W_\tau[\rho]$ (which
for this purpose is needed as a functional of the colour source $\rho_a$
in the covariant gauge).

\subsection{The Renormalization Group Equation at Small $x$}

When decreasing $x$ (or increasing the rapidity $\tau\equiv \ln(1/x)$),
new quantum modes become relatively ``fast'', and must
be included in the effective source at the new value of  $x$.
This operation modifies both the longitudinal support of $\rho$
(with increasing $\tau$, the colour source extends to larger and larger
values of the $x^-$), and its correlations. All these changes can be
absorbed into a functional renormalization of the weight function
$W_\tau[\rho]$,
which therefore depends upon $\tau$. To the order of interest, this
dependence is governed by a functional
{\it renormalization group equation}, which reads \cite{JKLW97,PI} :
\be\label{RGE}
{\del W_\tau[\rho] \over {\del \tau}}\,=\,
 {1\over 2} \int_{x_\perp,y_\perp}\,{\delta \over {\delta
\rho_\tau^a(x_\perp)}}\,\chi_{ab}(x_\perp, y_\perp)[\rho]\, 
{\delta \over \delta \rho_\tau^b(y_\perp)}\,W_\tau[\rho]\,,\ee
with $\chi[\rho]$ a positive definite functional of $\rho$, 
whose explicit form is not needed here (this can be found in Refs.
\cite{PI,SAT,Cargese}), but whose relevant properties will be discussed below.

Eq.~(\ref{RGE}) allows one to construct the weight 
function by integrating out the quantum gluons in layers of $\tau$.
The kernel $\chi[\rho]$ encodes the changes in the correlations
of the colour source induced at one step in this construction, 
and depends upon the source $\rho$ created in all the previous steps.
Indeed, the  quantum gluons which are integrated out in one step
propagate through the background field $A[\rho]$ of the
colour source generated in the previous steps. In general, this
classical field is strong, so its non--linear effects on the
dynamics of the quantum gluons must be included exactly.
Because of that,  the functional $\chi[\rho]$ is non--linear in  $\rho$ to
all orders, and also non--local, both in the longitudinal
coordinates, and in the transverse ones.
In fact, the non--linearity in  $\rho$ and the non--locality in $x^-$
are strongly correlated, since they have a common origin: namely, the
fact that $\chi[\rho]$ depends upon $\rho$ via the Wilson lines
in eq.~(\ref{v}).

There is one more notation in eq.~(\ref{RGE}) which waits for
an explanation: this is the argument $\rho_\tau^a(x_\perp)$ of the 
functional derivatives there. This is a notation for the colour
charge density  $\rho^a(x^-,x_\perp)$ at a specific longitudinal
location $x^-=x^-_\tau$, which is the upper boundary of the
support of the classical source in $x^-$. That is, the
colour source at rapidity $\tau$ has support within a
limited interval in $x^-$, namely at
$0 < x^- < x^-_\tau$, with $x^-_\tau\propto {\rm e}^{\tau}$.

This correlation between the quantum evolution in $\tau$ and the
longitudinal extent of the colour charge distribution
can be understood as a consequence of the uncertainty principle:
Since obtained by integrating out 
quantum modes with large longitudinal momenta $p^+ \gg xP^+\equiv
{\rm e}^{-\tau}P^+$, the classical source at rapidity $\tau$
must be localized near $x^-=0$, within a distance 
$\Delta x^-\sim  {\rm e}^{\tau}x^-_0$ (with $x^-_0\equiv 1/P^+$).

But eq.~(\ref{RGE}) shows that this correlation is even stronger:
Namely, it shows that, when the rapidity is further increased, say from $\tau$ 
to $\tau+{\rm d}\tau$, the additional contribution to the colour source
which is generated in this way has {\it no overlap} in $x^-$ with
the original source at rapidity $\tau$. Rather, this new  contribution
makes a new layer in $x^-$, which is located
between $ x^-_\tau$ and  $ x^-_{\tau+d\tau}$.
This is why the functional derivatives in  eq.~(\ref{RGE}) 
involve just the colour source $\rho_\tau\equiv 
\rho(x^-_\tau)$ in this outermost layer.

To conclude, when integrating out the quantum modes in 
layers of $\tau$, one builds the colour source in layers of $x^-$, 
with a one--to--one correspondence between 
the $x^-$ coordinate of a given layer in $\rho$ and the rapidity $\tau$
of the modes that have been integrated out to generate that layer. 

This correspondence is most simply formulated if one uses
 the {\it space--time rapidity} y,
\be {\rm y}\,\equiv\, \ln(x^-/ x^-_0),\quad 
x^-_0\equiv 1/P^+\,,\quad -\infty < {\rm y} < \infty\,,\ee
to indicate the longitudinal
coordinate of a field. We shall set, e.g.,
\be\label{rhotau}
\rho_{{\rm y}}^a(x_\perp)&\equiv&
 x^-\rho^a(x^-,x_\perp)\qquad {\rm for}\quad
x^-=x^-_{\rm y}\equiv x^-_0{\rm e}^{{\rm y}},\nn
\int d{\rm y}\,\rho_{{\rm y}}^a(x_\perp)&=&\int dx^-\,\rho^a(x^-,x_\perp),
\ee
and similarly for $\alpha$, eq.~(\ref{alpha}), or any other field.
Eq.~(\ref{v}) is then rewritten as :
\be\label{vy}
V^\dagger(x_{\perp})\,=\,{\rm P} \exp
 \left \{ig \!\int\! d{\rm y}\,\alpha^a_{\rm y} (x_{\perp})t^a
 \right \}.\ee

Then, the previous discussion shows that the space--time rapidity y
of a given layer in $\rho$ is identical to the usual (momentum) rapidity
of the fast gluons that have produced that layer. In particular, the
colour source created by the quantum evolution up to $\tau$ has support
at space-time rapidities ${\rm y}\le \tau$ (in agreement with the simple
argument based on the uncertainty principle). This is important since,
as we shall see, the  $\tau$--dependence of the observables computed
in the effective theory comes precisely from the upper limit on the
longitudinal support of $\rho$. The Colour Glass evolves by expanding
in y.

\section{The Gaussian approximation}
\setcounter{equation}{0}

The exact solution to eq.~(\ref{RGE}) is certainly
not a Gaussian. The kernel $\chi[\rho]$ in this equation is 
non-linear in $\rho$ to all orders,
so the complete solution $W_\tau[\rho]$ must be non-linear as well,
for any initial condition.
 (This non-linearity is  manifest on the 
formal representation of the solution derived in Ref. \cite{PATH}.)
Still, as mentioned in the Introduction,
there are interesting situations in which a weight function
which is a Gaussian in $\rho$ may capture the relevant physics. 
Before describing such physical situations in more detail, in
Sections 4 and 5, let us first characterize the general structure
of a Gaussian weight function,  and show how to obtain such a Gaussian
approximation for $W_{\tau}[\rho]$, at least at a formal level, via
{ mean field approximations} to the non--linear RGE (\ref{RGE}) 
(see also Ref. \cite{SAT}).

We start by displaying the most general Gaussian 
weight function which is consistent with gauge symmetry,
and respects the correspondence between
space--time rapidity and momentum rapidity explained in the previous
section. In the covariant
gauge, in which both the classical solution and the kernel
$\chi[\rho]$ are known explicitly, 
this Gaussian reads:
\be\label{Wgauss}
{W}_{\tau}[\rho_{\rm y}^a(x_{\perp})]
\,=\,{\cal N}_\tau\,
{\rm exp}\!\left\{-\,{1 \over 2}
\int_{-\infty}^\tau d{\rm y}
\int_{x_\perp,y_\perp}\!\frac{\rho_{\rm y}^a(x_{\perp})
\rho_{\rm y}^a(y_{\perp})}{\lambda_{\rm y}(x_{\perp},y_\perp)}
\right\}\,\,\times\,
\delta_\tau[\rho]\,,\ee
where $\delta_\tau[\rho]$ is a $\delta$--functional enforcing that
$\rho_{\rm y}\equiv 0$ for any ${\rm y}> \tau$, 
and ${\cal N}_\tau$ is a numerical factor,
which  ensures the right normalization:
\be\label{norm}
\int { D}\rho\, \,W_\tau[\rho]\,=\,1\,,\ee
but is irrelevant for the computation of correlation functions.

Note that the information about the longitudinal support of
$\rho_{\rm y}$ is included in the weight function. Thus, when
computing observables by averaging over $\rho$ with this
 weight function, like in eq.~(\ref{Ntau}), the 
support of $\rho_{\rm y}$ will be restricted to ${\rm y}\le \tau$,
as it should.

For physical space--time rapidities ${\rm y}\le \tau$,
eq.~(\ref{Wgauss}) is indeed a Gaussian in $\rho$, with the
following 2--point function :
\be\label{rhoHM}
\langle \rho_{\rm y}^{a}(x_\perp)\,
\rho_{{\rm y}'}^{b}(y_\perp)\rangle_\tau=\delta^{ab}\delta({\rm y}
-{\rm y}')\theta(\tau-{\rm y})\,\lambda_{\rm y}(x_\perp,y_\perp),
\ee
which is local in space--time rapidity, 
but generally non--local in the transverse coordinates.
This is the most general non--local structure which is allowed by
gauge symmetry, as we explain now: For $W_\tau[\rho]$ to be gauge--invariant,
 the colour sources $\rho_{\rm y}^{a}(x_\perp)$
at different points $({\rm y},x_\perp)$ must
be connected by appropriate gauge links, or ``Wilson lines''.
In the covariant gauge, in which eq.~(\ref{Wgauss}) is explicitly written,
the gauge potential has just a plus component,  $A^+_a \equiv \alpha_a$,
so the only non--trivial gauge links are those % Wilson lines
in the longitudinal direction, cf. eq.~(\ref{v}). And, indeed, the 
kernel $\chi[\rho]$ of the RGE is built with such Wilson lines \cite{PI,Cargese},
and thus is consistent with gauge symmetry, but also non--linear
in $\rho$ to all orders. For the gauge--invariance 
to be preserved in the Gaussian approximation, where no Wilson
lines are permitted, $W_\tau[\rho]$ must be {\it local} in y,
as shown in eq.~(\ref{Wgauss}).

The Gaussian weight function (\ref{Wgauss}) 
obeys the following evolution equation:
\be\label{RGE-MFA}
{\del  W_\tau[\rho] \over {\del \tau}}\,=\,{1 \over 2}
\int_{x_\perp,y_\perp}\!\lambda_\tau(x_{\perp},y_\perp)\,
{\delta^2  W_\tau[\rho]\over {\delta\rho_\tau^a(x_{\perp})
\delta \rho_\tau^a(y_{\perp})}}\,.
\ee
By comparison with eq.~(\ref{RGE}), one sees that, at a formal level, 
the Gaussian approximation is tantamount to replacing the general
kernel $\chi[\rho]$ in the RGE by the $\rho$--independent, but $\tau$--dependent,
quantity $\lambda_\tau$. 
%The precise meaning of such a replacement
%turns out to be quite subtle, and will be clarified in what follows.

To determine the kernel $\lambda_\tau$ of the Gaussian,
which is also the 2--point function in the present
approximation, we shall require this quantity 
to obey the ``mean field version'' of the actual
evolution equation  satisfied by the complete 2--point function 
$\langle \rho(1)\rho(2)\rangle_\tau$. Here, by ``mean field version'',
we mean that the correlations which enter this equation are 
evaluated with the 
Gaussian weight function (\ref{Wgauss}) itself. This gives
a {self-consistent} approximation for the  2--point function.

Since the general 2--point function 
$\langle \rho_{\rm y}^{a}(x_\perp)\,
\rho_{{\rm y}'}^{b}(y_\perp)\rangle_\tau$ is not local in y, nor diagonal
in colour, it is preferable to consider the equation satisfied by the
following quantity:
 \be\label{mutau}
\mu_\tau(x_\perp,y_\perp)\,\equiv \,\frac{1}{N_c^2-1}\,
\langle \rho^{a}(x_\perp)\,\rho^{a}(y_\perp)\rangle_\tau,\ee
where
\be\label{rhoT}
\rho^a(x_{\perp})\,\equiv\, \int d{\rm y}\,\rho_{\rm y}^a (x_{\perp})\ee
is the effective colour charge density in the transverse plane, as
obtained after integrating over the longitudinal profile of the hadron.

In the Gaussian approximation, eq.~(\ref{rhoHM}) immediately implies:
\be\label{mutauMFA}
\mu_\tau(x_\perp,y_\perp)\,=\, \int_{-\infty}^\tau  d{\rm y}\,
\lambda_{\rm y}(x_{\perp},y_\perp)\,,\qquad {\rm or}\qquad
\lambda_\tau(x_{\perp},y_\perp)\,=\,{\del  \mu_\tau
\over {\del \tau}}(x_\perp,y_\perp).\ee

In general, the evolution equation for
the two-point function (\ref{mutau})  can be obtained from its definition:
\be\label{rho2}
\langle \rho^{a}(x_\perp)\,
\rho^{b}(y_\perp)\rangle_\tau\,\equiv\,
\int { D}\rho\, \,W_\tau[\rho]\,\, \rho^{a}(x_\perp)\,
\rho^{b}(y_\perp),\ee
by first taking a derivative with respect to $\tau$, then using
eq.~(\ref{RGE}) for ${\del  W_\tau / {\del \tau}}$, and finally  performing
a couple of functional integrations by parts. This yields:
\be\label{rho2ev}
\hspace*{-.5cm}
{\del  \over {\del \tau}}\,\langle \rho^{a}(x_\perp)\,
\rho^{b}(y_\perp)\rangle_\tau \,=\,\langle \chi^{ab}(x_\perp,y_\perp)
\rangle_\tau \,+\,\langle\sigma^{a}(x_\perp)\,\rho^{b}(y_\perp)\rangle_\tau \,
+\,\langle \rho^{a}(x_\perp)\,\sigma^{b}(y_\perp)\rangle_\tau\,,\ee
where the quantity $\sigma^a(x_\perp)$ has been generated as the functional
 derivative of $\chi[\rho]$:
\be\label{sigma}
\sigma^a(x_\perp)[\rho]\,\equiv\,\frac{1}{2}\,\int  d^2y_\perp\, 
{\delta \chi^{ab}(x_\perp, y_\perp)\over 
\delta\rho_\tau^b(y_\perp)}\,.\ee
It is useful to recall here that, 
in the quantum calculation leading to eq.~(\ref{rhoHM})
\cite{JKLW97,PI}, $\sigma[\rho]$ represents the ``virtual correction'' 
(i.e., the sum of the self-energy and vertex corrections), 
and is complementary to the ``real correction''  represented by  $\chi[\rho]$. 
The relation (\ref{sigma}) between these
two quantities \cite{W,PI} is very important, since it ensures that
the evolution equations for gauge-invariant observables 
derived from  eq.~(\ref{RGE}) are infrared finite.

In general, both $\chi$ and $\sigma$ are non-linear in $\rho$, so
eq.~(\ref{rho2ev}) is not a closed equation --- it couples the
2--point function to higher $n$--point functions of $\rho$ ---,
but only the first in an infinite hierarchy of coupled evolution equations.
A closed equation can be
nevertheless obtained in the {\it mean field approximation} (MFA) in which
the correlation functions in eq.~(\ref{rho2ev}) are evaluated with the Gaussian
weight function (\ref{Wgauss}). Then, 
eq.~(\ref{rho2ev}) reduces to the
following equation for $\lambda_\tau = {\del  \mu_\tau / {\del \tau}}$
(cf. eq.~(\ref{mutauMFA})) :
\be\label{lambdaeq}
\lambda_\tau(x_{\perp},y_\perp)\,=\,\langle \chi(x_\perp,y_\perp)\,+\,
%\rangle_\tau \,+\,\langle\sigma(x_\perp)\,\rho(y_\perp)\rangle_\tau \,
\sigma(x_\perp)\,\rho(y_\perp) \,+\, 
\rho(x_\perp)\,\sigma(y_\perp)\rangle_\tau\,,\ee
where $\langle \chi\rangle\equiv %\frac{1}{N_c^2-1} 
\langle \chi^{aa} \rangle/(N_c^2-1)$,
etc., and the expectation value in the r.h.s. is computed with the
weight function (\ref{Wgauss}), so it is a known functional of
$\lambda_{\rm y}$. In general, this functional is very 
non-linear and also non-local, but it considerably  simplifies
in the limiting cases of interest (see Sect. 5 below).

To summarize, our MFA is defined by the weight function
(\ref{Wgauss}) with the kernel $\lambda_\tau$ determined 
self-consistently by eq.~(\ref{lambdaeq}). 
This is more general than the MFA considered
in Ref. \cite{SAT}, which was an approximation
on the evolution Hamiltonian (the functional
differential operator in the r.h.s. of eq.~(\ref{RGE})) :
\be\label{HMFA}
H\,\equiv \,-\,
{1\over 2} \,{ \delta \over {\delta
\rho_\tau}}\,\chi[\rho]\, 
{ \delta \over \delta \rho_\tau}\,\,\longrightarrow\,\, {\bar H}
\,\equiv \,-\,{1\over 2} \,\langle \chi[\rho]\rangle_\tau\,
{\delta^2  \over {\delta\rho_\tau \delta \rho_\tau}}\,.\ee
In eq.~(\ref{HMFA}), the kernel $\chi[\rho]$ in the  Hamiltonian
 is simply replaced by
its average over $\rho$, so its functional derivative $\sigma[\rho]$
%eq.~(\ref{sigma}), 
is implicitly neglected, and eq.~(\ref{lambdaeq}) reduces to 
\be\label{gammatau}
\lambda_\tau(x_{\perp},y_\perp)\,=\,
\langle \chi(x_\perp,y_\perp)\rangle_\tau\,,\ee
as is also obvious by comparing eqs.~(\ref{RGE-MFA}) and (\ref{HMFA}).
The reason why eq.~(\ref{lambdaeq}) appears to be more complete,
is that, in its derivation, the 
equation for the 2--point function has been first generated 
with the {\it full} Hamiltonian,
and the MFA has been implemented only afterwards. By contrast,
eq.~(\ref{gammatau}) has been generated directly with an approximate
form of the  Hamiltonian.

As we shall discuss in Sect. 5, the difference 
between eqs.~(\ref{lambdaeq}) and (\ref{gammatau}) has important consequences.
In particular, eq.~(\ref{gammatau}) is not good enough for our
present purposes. Rather, our final approximation for the weight function,
to be constructed in Sect. 6, will be based on  eq.~(\ref{lambdaeq}).

We conclude this section by displaying the expression for
the dipole-hadron scattering amplitude (\ref{Ntau}) in the Gaussian 
approximation. The Wilson lines in eq.~(\ref{vy}) 
involve the Coulomb field %covariant-gauge 
$\alpha^a_{\rm y}(x_{\perp})$, which is linearly related to the colour
charge density (cf. eq.~(\ref{Poisson})), and therefore
is itself a Gaussian random variable, with 2--point function 
(cf eq.~(\ref{rhoHM})) :
\be\label{alphaHM}
\langle \alpha_{\rm y}^{a}(x_\perp)\,
\alpha_{{\rm y}'}^{b}(y_\perp)\rangle_\tau&=&\delta^{ab}\delta({\rm y}
-{\rm y}')\theta(\tau-{\rm y})\,\gamma_{\rm y}(x_\perp,y_\perp),\ee
where %ith ``propagator'' 
%$\gamma_{\rm y}$ related to the charge-charge correlator 
%$\lambda_{\rm y}$ via
\be
\gamma_{\rm y}(x_\perp,y_\perp)\equiv \int d^2z_\perp d^2u_\perp
\langle x_\perp|\frac{1}{-\grad^2_\perp}|z_\perp\rangle\,
\lambda_{\rm y}(z_\perp,u_\perp)\,
\langle u_\perp|\frac{1}{-\grad^2_\perp}|y_\perp\rangle.
\label{gamlam}\ee
From eqs.~(\ref{vy}) and (\ref{alphaHM}), a straightforward
calculation yields \cite{JKMW97,Cargese} :
\be\label{Stau-MFA}
S_\tau(x_\perp,y_\perp)={\rm exp}\left\{
-{g^2C_R \over 2}\int_{-\infty}^\tau d{\rm y}
\Big[\gamma_{\rm y}(x_{\perp},x_\perp)
+ \gamma_{\rm y}(y_{\perp},y_\perp) - 
2\gamma_{\rm y}(x_{\perp},y_\perp)\Big]\right\}.\ee
The colour factor $C_R$ is $C_F=(N_c^2-1)/2N_c$ for a dipole made of
a quark and an antiquark (this is the relevant case for $\gamma^* p$
deep inelastic scattering), and $C_A=N_c$ for a gluonic dipole,
which is interesting for the discussion of the RGE
(since the Wilson lines which enter the kernel $\chi[\rho]$ 
are written in the adjoint representation, i.e., they refer to gluons).

Eq.~(\ref{Stau-MFA}) simplifies if one assumes the hadron to be homogeneous 
in the transverse plane, within a disk of radius $R$, with $R$
much larger than any transverse separation $x_\perp-y_\perp$ of interest
(so that one can neglect the edge effects).  Then, the various
2--point functions depend only upon the relative coordinate: e.g.,
$\lambda_{\rm y}(x_\perp,y_\perp)=\lambda_{\rm y}(r_\perp)$,
with $r_\perp\equiv x_\perp-y_\perp$, and it is convenient
to introduce the Fourier transform:
\be\label{lamk}
\lambda_{\rm y}(k_\perp)\,=\,\int d^2r_\perp {\rm e}^{-i k_\perp
\cdot r_\perp}\,\lambda_{\rm y}(r_\perp),\ee
and similarly for the other functions. The relation (\ref{gamlam})
simplifies to:
\be\label{gammak}
 \gamma_\tau(k_\perp)\,=\,\frac{\lambda_{\rm y}(k_\perp)}{k_\perp^4}\,,\ee
while eq.~(\ref{Stau-MFA}) is finally rewritten as:
\be\label{Stau-MOM}
S_\tau(r_\perp)={\rm exp}\left\{-g^2C_R\int_{-\infty}^\tau d{\rm y}
\int \!{d^2k_\perp\over (2\pi)^2}\,\frac{\lambda_{\rm y}(k_\perp)}{k_\perp^4}\,
\Bigl[1-
{\rm e}^{ik_\perp\cdot r_\perp}\Bigr]\right\}.\ee

\section{The McLerran-Venugopalan model}
\setcounter{equation}{0}

In this section, we interrupt our presentation of the quantum
evolution and briefly review a simple Gaussian model for the
weight function, originally proposed by McLerran and Venugopalan 
\cite{MV} to describe the gluon distribution of a large nucleus ($A\gg 1$)
(see also \cite{K96,JKMW97,KM98,LM00,Cargese}).
There is no quantum evolution in this model (i.e., no dependence upon $x$),
but this can be viewed as the initial condition, valid at
some moderate value $x_0$, for the quantum evolution towards smaller
values of $x$.

In this model, the only colour sources are the
$A\times N_c$ valence quarks, which are assumed to be uncorrelated,
since they do not overlap with each other:
Valence quarks which overlap in transverse projection 
belong typically to different nucleons, 
and thus have no overlap in the longitudinal direction,
because of confinement. 
%(We assume here, of course, that the transverse momenta are large
%enough to probe the individual constituents of a nucleon, and not
%the nucleon as a whole.) 
The distribution of these sources is therefore
described by a Gaussian weight function, 
like in eq.~(\ref{Wgauss}), but with a kernel which is local also
in the transverse coordinates:
\be\label{WMV}
{W}_{\tau_0}[\rho]
\,=\,{\cal N}_\tau\,
{\rm exp}\!\left\{-\,{1 \over 2}
\int_{-\infty}^{\tau_0} d{\rm y}
\int_{x_\perp}\!\frac{\rho_{\rm y}^a(x_{\perp})
\rho_{\rm y}^a(x_{\perp})}{\lambda_{\rm y}(x_{\perp})}
\right\}\,.\ee
Physically, 
the kernel $\lambda_{\rm y}(x_{\perp})$ represents the average colour 
charge squared of the valence quarks per unit space-time rapidity and
per unit transverse area, at the given point $({\rm y},x_{\perp})$.
It satisfies the sum rule:
\be
\int d{\rm y} \int d^2{x_\perp}\,\lambda_{\rm y}(x_{\perp})\,=\,
\frac{g^2 C_F A N_c}{N_c^2-1}\,=\,\frac{g^2 A}{2}\,,\ee
which follows since the total colour squared in the nucleus is 
$\langle {\cal Q}^a {\cal Q}^a \rangle= g^2 C_F A N_c$ 
(recall that the colour charge squared 
of a single quark is $g^2t^at^a=g^2C_F$), with
\be\label{Qtot}
{\cal Q}^a \,\equiv\,\int d{\rm y} \int d^2{x_\perp}\,
\rho_{\rm y}^a(x_{\perp}).\ee
The y--dependence of the kernel $\lambda_{\rm y}(x_{\perp})$ remains
unspecified in the MV model. This is not necessarily a problem since,
when computed in this model, many interesting quantities
(see, e.g., eq.~(\ref{Stau-MOM}))
involve only the colour charge integrated over y
(cf. eqs.~(\ref{mutau})--(\ref{mutauMFA})):
\be\label{mu0}
\mu_{0}(x_{\perp})\,\equiv\,\int_{-\infty}^{\tau_0} d{\rm y}\,
\lambda_{\rm y}(x_{\perp})\,.\ee
To get simple formulae, it is again convenient to assume
transverse homogeneity within the nuclear disk 
of radius $R_A= R_0 A^{1/3}$ (this assumption can be relaxed, 
and actually is, in numerical simulations \cite{KV}). Then:
\be
\mu_{0}(A)\,=\,\frac{g^2 A}{2\pi R^2_A}\,\propto\,A^{1/3}.
\ee

When the MV model is used to compute physical quantities, one usually
finds logarithmic infrared divergences, which can be
attributed to the fact that the Gaussian weight function (\ref{WMV}) 
is local in $x_{\perp}$. Note that, when translated to momentum space,
this locality means that the Fourier modes of $\lambda_{\rm y}$ are
independent of $\kk$ (cf. eq.~(\ref{lamk})) :
\be\label{lamMV} \lambda_{\rm y}(x_\perp,y_\perp)=
\delta^{(2)}(x_\perp-y_\perp)\lambda_{\rm y}\,\,\,\,\Longrightarrow
\,\,\,\,\lambda_{\rm y}(\kk)\,=\,\lambda_{\rm y}\,.\ee
Consider then eq.~(\ref{Stau-MOM}) for the dipole scattering amplitude.
In the MV model, it becomes:
\be\label{SMV}
S_0(r_\perp)={\rm exp}\left\{-g^2C_R
\int \!{d^2k_\perp\over (2\pi)^2}\,\frac{\mu_0}{k_\perp^4}\,
\Bigl[1-
{\rm e}^{ik_\perp\cdot r_\perp}\Bigr]\right\},\ee
and the integral over $k_\perp$ in the exponent has a 
logarithmic infrared divergence indeed\footnote{The divergence
is only logarithmic since, at small $k_\perp$,
$1-{\rm e}^{ik_\perp\cdot r_\perp}\approx - ik_\perp\cdot r_\perp
+ (1/2)(k_\perp\cdot r_\perp)^2$, and the term linear in $k_\perp$
vanishes after angular integration. This compensation of the leading 
IR divergence between 1 and ${\rm e}^{ik_\perp\cdot r_\perp}$
is ultimately a consequence of the fact that the external probe 
is a colourless {\it dipole}, with a gauge--invariant $S$-matrix element.}.

Clearly, this divergence reflects the limitations of the assumptions
underlying the MV model. In particular, the assumption that the
colour sources are uncorrelated must fail for sufficiently
large transverse separations. For instance, long--range correlations
will certainly occur because of confinement:
The three valence quarks within the same nucleon
are bound in a colour singlet state, so the total colour charge 
(\ref{Qtot}), together with its higher multipolar moments, must vanish
when measured over distances of the order of the  nucleon size
$R_0$, or larger. This in turn requires the correlator
$\lambda_{\rm y}(\kk)$ to vanish at low momenta
$\kk\simle 1/R_0$, in contradiction with eq.~(\ref{lamMV}).

Based on this physical idea, it has been proposed \cite{LM00} to
cure the infrared problem of the MV model by imposing colour neutrality 
on the nucleon level. This means that, effectively, the integral
over $\kk$ in eq.~(\ref{SMV}) is cut off in the infrared at momenta
of order $1/R_0\sim \Lambda_{QCD}$. In fact, since the divergence
is only logarithmic, then, to leading--log accuracy,
%i.e., in an approximation
%which retains only the large logarithm $\ln(1/r_\perp\Lambda_{QCD})$,
the exact value of the cutoff is not important, neither are the details
of the non--perturbative mechanism responsible for this screening.
To evaluate eq.~(\ref{SMV}) to this accuracy,
one can insert a sharp momentum cutoff equal to $\Lambda_{QCD}$,
and also expand the exponential (since $k_\perp\cdot r_\perp\ll 1$
for this dominant contribution). One obtains
\cite{JKMW97,Cargese} :
\be\label{SMV1}
S_0(r_\perp)\,\simeq\,
{\rm exp}\left\{-\,\frac{r_\perp^2 Q_0^2(A)}{4}\, %\alpha_s C_F \,\mu_0\,
\ln{1\over r_\perp^2\Lambda^2_{QCD}}\right\},\ee
with:
\be\label{SMV2}
Q_0^2(A)\,\equiv\,\alpha_s C_R \,\mu_0(A)\,=\,\frac{2\alpha_s^2 C_R
A^{1/3}}{R_0^2}\,.\ee
Still, for the physical scales of interest (e.g., in relation with
heavy ion collisions), the leading--log approximation above is not
really satisfactory, and the complete, numerical results in Ref. \cite{KV}
turn out to be quite sensitive to the specific assumptions about the
non--perturbative physics (i.e., upon the prescription
used to implement colour neutrality).

Thus, in this classical context, where the
valence quarks are the only colour sources, 
one cannot avoid a rather strong sensitivity to the soft,
non--perturbative physics. As we shall show in what follows,
this sensitivity disappears once the effects of the % non--linear
quantum evolution are taken into account: At 
sufficiently large energies (or small enough values of $x$), 
the colour sources  are predominantly {\it gluons}, and these
gluons have long--range correlations associated with the non--linear
(but still perturbative) physics of saturation,  which ensure
colour neutrality already over the short scale $1/Q_s
\ll 1/\Lambda_{QCD}$.

\section{Colour neutrality from quantum evolution}
\setcounter{equation}{0}

We now return to the quantum evolution described by eq.~(\ref{RGE}),
and examine the validity of the Gaussian approximations obtained
in Sect. 3 via formal manipulations.
To this aim, we need to consider the structure of 
the kernel $\chi[\rho]$ in some detail. In fact, for the present
purposes, it is enough to know that $\chi$ depends upon $\rho$ 
via the Wilson lines (\ref{vy}), and that the strength 
of the field $\alpha$ in these Wilson lines is correlated with 
the gluon density in the system (see Sect. 7.1). 
This suggests different approximations at {short} and,
respectively, {large} distances compared to the saturation length $1/Q_s$.

\subsection{High $\kk$ : The BFKL equation} 

For {short distances} $\rr\ll 1/Q_s(\tau)$, that is,
for an external probe with a large transverse resolution $Q^2=1/\rr^2$
(e.g., a small dipole with transverse size $\rr$),
scattering is dominated by gluons with relatively large transverse momenta
($\kk\gg Q_s(\tau)$), and thus a low density. In this {\it dilute} regime,
the colour field is weak,
$g\alpha\ll 1$, so one can expand the Wilson lines (\ref{vy})
in powers of $g\alpha$,
and keep only the linear term in the expansion:
\be\label{vlinear}
V^\dagger(x_{\perp})\,\approx\,1\,+\,
ig \!\int\! d{\rm y}\,\alpha_{\rm y} (x_{\perp})
\,\equiv\,1\,+\,ig\alpha(x_{\perp}).\ee
After this expansion, $\chi[\rho]$ becomes { quadratic} in
$\rho(x_{\perp})\equiv \int\! d{\rm y}\,\rho_{\rm y} (x_{\perp})$
(cf. eq.~(\ref{rhoT})) \cite{JKLW97,PI}:
\be\label{chilin}
\chi[\rho]\,\approx\,\rho\,{\cal K}\,\rho,\ee
%with a kernel ${\cal K}$ which is
%non-local in the transverse coordinates,
%and non-diagonal in colour  \cite{JKLW97,PI}. 
which means that the corresponding RGE is still non-linear :
\be\label{RGE-lin}
{\del  W_\tau[\rho] \over {\del \tau}}\,\approx\,{1 \over 2}\,
{\delta \over {\delta\rho}}\Big(\rho\,{\cal K}\,\rho\Big)
{\delta \over {\delta\rho}}\,W_\tau[\rho]\,.
\ee
That is, even 
in the {dilute} regime at high $\kk$, the weight function $W_\tau[\rho]$
is, strictly speaking, {\it not} a Gaussian. Still, as compared to the 
general RGE (\ref{RGE}), the evolution generated by
eq.~(\ref{RGE-lin}) exhibits an important simplification:
it does not mix correlations $\langle \rho(1)\rho(2)\dots\rho(n)\rangle$
with different numbers $n$ of colour sources \cite{PI}.

In particular, eq.~(\ref{RGE-lin}) provides a closed evolution equation 
for the 2--point function (\ref{mutau}) --- i.e., eq.~(\ref{rho2ev}) with
$\chi[\rho]\approx \rho\,{\cal K}\,\rho$ and 
$\sigma[\rho]\approx {\cal K}\,\rho$ ---, which is moreover {\it linear}
\cite{JKLW97}.  When written
in momentum space, this equation
is recognized  as the BFKL equation\cite{BFKL}:
\begin{eqnarray}
 {\partial \mu_\tau (k_{\perp}) \over \partial \tau} & = & \,\,\,
{\alpha_s N_c \over \pi^2}\,
\int d^2 p_{\perp}
 {k^2_{\perp} \over p^2_{\perp} (k_{\perp}-p_{\perp})^2}\,
 \mu_\tau(p_{\perp}) \nonumber \\
& & -\,
{\alpha_s N_c \over2 \pi^2}\,
\int d^2 p_{\perp}
 {k^2_{\perp} \over p^2_{\perp} (k_{\perp}-p_{\perp})^2}\,
 \mu_\tau(k_{\perp})\,.
\label{BFKL}
\end{eqnarray}
The emergence of the BFKL equation in this regime, and
in the present approximations, is as expected, since the quantity
$\mu_\tau (k_{\perp})$ plays also the role of
 the unintegrated gluon distribution (see Sect. 7.1).
The two terms in the r.h.s. are %of  eq.~(\ref{BFKL}) are
the ``real'' and, respectively, ``virtual'' BFKL corrections,
and originate from the terms involving $\chi$ and, respectively,  
$\sigma$, in the r.h.s. of eq.~(\ref{rho2ev}). Since  eq.~(\ref{BFKL})
is local in rapidity, then, clearly, if $\mu_\tau(\kk)$ satisfies this
equation, so does also  
$\lambda_\tau(\kk) = {\partial \mu_\tau/ \partial \tau}$, which
is the kernel of the Gaussian.

Let us now compare eq.~(\ref{BFKL}) with the corresponding predictions
of the mean field approximations introduced in Sect. 3, namely, 
eqs.~(\ref{lambdaeq}) and (\ref{gammatau}). 

In this low density regime, where $\chi$ is quadratic in $\rho$,
while $\sigma$ is linear, there is obviously no difference between the general
equation (\ref{rho2ev}) and its mean field approximation (\ref{lambdaeq}).
Thus, the BFKL equation (\ref{BFKL}) is {\it exactly}
reproduced by the linearized version of eq.~(\ref{lambdaeq}).
That is, although the general solution $W_\tau[\rho]$ to eq.~(\ref{RGE-lin}) 
is {\it not} a Gaussian, our mean field approximation is still
able to encode the evolution of the 2--point function correctly,
because this evolution is linear, and it 
does not mix  the 2--point function with higher $n$--point
functions (i.e., because the BFKL Hamiltonian in the r.h.s. of
eq.~(\ref{RGE-lin}) is diagonal in the number of fields $\rho$).

It is also important to notice that, within this Gaussian effective theory,
the BFKL evolution of the kernel $\lambda_\tau$ of the Gaussian 
gets transmitted to other 2--point functions of physical interest, like
the dipole--hadron scattering amplitude (\ref{Ntau}). Specifically, in the 
linear approximation appropriate at high $\kk$,  eq.~(\ref{Stau-MOM}) reduces to:
\be\label{Nlinear}
{\cal N}_\tau(r_\perp)\, =\, 1-S_\tau(r_\perp)\,\approx\,
g^2N_c \int_{-\infty}^\tau d{\rm y}
\int \!{d^2k_\perp\over (2\pi)^2}\,\frac{\lambda_{\rm y}(\kk)}{k_\perp^4}\,
%\mu_{\tau}(k_\perp)}{k_\perp^4}\,
\Bigl[1-
{\rm e}^{ik_\perp\cdot r_\perp}\Bigr].\ee
By taking a derivative with respect to $\tau$ in this equation, 
and using eq.~(\ref{BFKL}) 
for $\lambda_\tau = {\partial \mu_\tau/ \partial \tau}$, one can check, after
some lengthy but straightforward manipulations, that the (linearized) 
scattering amplitude obeys to:
\be\label{linearBK}
\frac{\del }{\del \tau} \, {\cal N}_\tau(\rr)\, =\, \bar\alpha_s \int 
{d^2z_\perp\over \pi}\, \, 
\frac{\rr^2}{(\rr-\z)^2\z^2}\, \left({\cal N}_\tau(\z)
-\frac12 \, {\cal N}_\tau(\rr)\right),\ee
which is the coordinate space form of the BFKL equation.
This is further recognized as the linearized version of the 
Balitsky--Kovchegov (BK) equation, which is the general evolution
equation for ${\cal N}_\tau(r_\perp)$ that follows from the RGE (\ref{RGE})
\cite{PI}, and also from other formalisms which focus on the evolution
of the dipole wavefunction \cite{B,K,Braun,W,AM01}.
%Note that, except in the large $N_c$ limit \cite{K,Braun}, this general
%equation is not a closed equation \cite{B,W,PI}.

Consider also the MFA based on eq.~(\ref{gammatau}) : 
Since this approximation misses the $\sigma$ term, it fails to 
reproduce the ``virtual''  BFKL term, and thus the  complete
BFKL equation (\ref{BFKL}). This is a serious failure:
as well known, the ``virtual'' term cancels the infrared divergence
of the ``real'' term at $p_{\perp}=k_{\perp}$, and thus plays an 
important role in the BFKL dynamics.

Still, there exists a kinematical limit in which the
``virtual'' term  becomes unimportant: this is
high--$k_{\perp}^2$ regime where one
retains only terms enhanced by both energy logs 
$\ln(1/x)$ and transverse momentum logs $\ln(k_{\perp}^2/\Lambda^2)$
(``double logarithmic approximation'', or DLA). This limit is common
to the BFKL  \cite{BFKL} and DGLAP  \cite{DGLAP} equations, and
is formally obtained  by assuming
 $p_{\perp}\ll k_{\perp}$ in eq.~(\ref{BFKL}).
In this limit, one expects the MFA based on eq.~(\ref{gammatau})
to become appropriate. This is confirmed by the analysis
in Ref. \cite{SAT} which shows that, at very high $\kk$,
the self-consistency equation (\ref{gammatau}) reduces to:
\be\label{DLA-lam}
\lambda_\tau(\kk)\,=\,\frac{\alpha_s N_c}{\pi} 
\int_{-\infty}^\tau d{\rm y} 
\int^{k_\perp^2}\frac{dp_\perp^2}{p_\perp^2}\,\lambda_{\rm y} 
(p_\perp),\ee
or, equivalently (cf. eq.~(\ref{mutauMFA})),
\be\label{DLA}
{\partial \mu_\tau(k_\perp)\over \partial \tau}\,=\,\frac{\alpha_s N_c}
{\pi} \int^{k_\perp^2}\frac{dp_\perp^2}{p_\perp^2}\,
\mu_\tau(p_\perp)\,,\ee
which is recognized indeed as the DLA evolution equation
(compare to  eq.~(\ref{BFKL})).

It is important to keep in mind, however, that DLA is not the right 
approximation 
to study the approach towards saturation. Indeed, there is a large
kinematical gap between the validity range for this approximation 
and the saturation line $\kk=Q_s(\tau)$ \cite{SCALING}, and,
within this gap, one can rely only on the full BFKL equation.
This is why in what follows we shall focus exclusively
(in this high momentum regime) on the BFKL dynamics.

\subsection{Low $\kk$ : Saturation and colour neutrality}

At low transverse momenta $\kk\simle Q_s$, the gluons are {\it saturated}, 
that is, they have large occupation numbers 
$dN/d\tau d^2\kk \sim 1/\alpha_s$, and a radial
momentum distribution $dN/d\tau d\kk$ which is peaked at
%$ \propto \kk d^3N/d^2\kk$ which is peaked at
the relatively hard scale $\kk\sim Q_s(\tau)$ (see Sect. 7.1). 
Thus, in the corresponding
CGC description, the colour fields have large amplitudes\footnote{More
precisely, we mean here the fields integrated over all
rapidities, i.e., $\alpha(x_{\perp}) =
\!\int\! d{\rm y} \alpha_{\rm y} (x_{\perp})$.}
$\alpha \sim 1/g$, and carry typical momenta  $\kk\sim Q_s$.
Accordingly, the Wilson lines (\ref{vy}) --- which are complex
exponentials built with these fields ---  oscillate around zero
over a characteristic distance $\sim 1/Q_s(\tau)$ in the transverse
plane. This implies that  Wilson lines which are separated by large 
distances $\gg 1/Q_s(\tau)$ are necessarily uncorrelated (since their
relative phases are random). 
Thus, when studying
the dynamics over large transverse separations 
$\rr\gg 1/Q_s(\tau)$, it should be a good approximation
to neglect the correlations of the Wilson lines
(or, more generally, to treat them as small quantities).
This is the ``random phase approximation'' (RPA) introduced in
Ref. \cite{SAT,W}.

In this approximation, the RGE (\ref{RGE}) simplifies drastically
\cite{SAT} : After neglecting the Wilson lines,
the kernel $\chi$ becomes independent of $\rho$, and the 
corresponding solution
%\footnote{Recall indeed
%that $\chi$ depends upon $\rho$ only via the Wilson lines \cite{PI}.}
$W_\tau[\rho]$ is truly a Gaussian (to the accuracy of the RPA).
This is quite remarkable, given that, at saturation, we are in
a strong field regime, where the dynamics is fully non--linear.
Specifically, one finds \cite{SAT} 
\be\label{RPA}
\chi^{ab}(x_\perp,y_\perp)\,\approx\,\delta^{ab}\,\frac{1}{\pi}\,
\langle x_\perp|{-\grad^2_\perp}|y_\perp\rangle\,\qquad ({\rm RPA})
\ee
which gives the following RGE :
\be\label{RGE-RPA}
{\del  W_\tau[\rho] \over {\del \tau}}\,=\,{1 \over 2}
\int\!\frac{d^2k_\perp}{(2\pi)^2}\,\,\frac{\kk^2}{\pi}\,
{\delta^2  W_\tau[\rho]\over {\delta\rho_\tau^a(k_{\perp})
\delta \rho_\tau^a(-k_{\perp})}}\,,
\ee
where we recognize a Coulomb--like Hamiltonian in the r.h.s. A brief comparison
with eq.~(\ref{RGE-MFA}) allows us to identify the charge-charge correlator
at low $\kk$ :
\be
\lambda_{\rm y}(\kk)&=&\frac{1}{\pi}\,\kk^2\,,\qquad (\kk\ll Q_s(\tau)).
\label{lamRPA}\ee
Clearly, both mean field approximations discussed in Sect. 3 reproduce 
eqs.~(\ref{RPA})--(\ref{lamRPA}) in the low--$\kk$ limit.

Note that this low--$\kk$ distribution is 
homogeneous in all the (longitudinal and transverse) coordinates.
That is, in the transverse plane, it is only a function of the
relative coordinate $x_\perp-y_\perp$, and in the longitudinal direction,
it is independent of the space-time rapidity y. One should
however keep in mind that,
for given $\kk\ll Q_s(\tau)$, eq.~(\ref{lamRPA}) applies
only for y in the interval $\tau_s(\kk)<{\rm y} < \tau$,
with $\tau_s(\kk)$ being the
rapidity at which the saturation scale becomes equal to the momentum
$\kk$ of interest:
\be\label{taus} 
Q_s^2(\tau)\,=\, \kk^2\qquad {\rm  for}\qquad \tau\,=\,\tau_s(\kk).
\ee
It follows that the integrated quantity (cf. eq.~(\ref{mutauMFA})) :
\be\label{mu-sat}
\mu_\tau(k_\perp)\Big |_{\rm sat}\,=\,
\int\limits_{\tau_s(k_\perp)}^\tau \! d{\rm y \,} \,\frac{k_\perp^2}{\pi}\,
= \Big(\tau-\tau_s(k_\perp)\Big)\frac{k_\perp^2}{\pi}\,,
\,\qquad (\kk\ll Q_s(\tau)),\ee
which measures the density of saturated colour sources (with given $k_\perp$)
in the transverse plane,
grows only {\it linearly} with $\tau$, that is, logarithmically with
the energy. This should be contrasted with the rapid
increase of the corresponding quantity at $\kk\gg Q_s(\tau)$,
which evolves according to the BFKL equation (\ref{BFKL}), and thus rises
exponentially with $\tau$. Eq.~(\ref{mu-sat}) shows that, at low momenta 
$k_\perp\ll Q_s(\tau)$, the colour charge density {\it saturates}, 
because of the strong non--linear effects in the quantum evolution \cite{SAT}. 
As we shall discuss in Sect. 7.1, this further implies the saturation of the
gluon distribution. % generated by these colour sources.
In Ref. \cite{AM02}, the result in eq.~(\ref{mu-sat}) has been
reobtained from a study of the BK equation.

According to eq.~(\ref{lamRPA}), the only remaining correlations at saturation
are those in the transverse separation $r_\perp$ between
the colour sources. Importantly,
these are such as to ensure {\it colour neutrality} \cite{Cargese,FB,AM02}.
Indeed, the fact that $\lambda_{\rm y}(\kk) \propto \kk^2$ when
$\kk^2\to 0$ implies $\langle {\cal Q}^a{\cal Q}^a \rangle=0$,
where ${\cal Q}^a$ is the total colour charge (\ref{Qtot}) of
the saturated gluons:
\be
\langle {\cal Q}^a{\cal Q}^a \rangle_\tau\,=\,\int d{\rm y} 
%_{\tau_s(\kk)}^\tau d{\rm y} 
\lambda_{\rm y}(\kk=0)\,=\,0.\ee
Since the distribution of $\rho$ at small $\kk$ is a Gaussian, 
and both the 1--point function $\langle {\cal Q}^a \rangle$, and the
2--point function $\langle {\cal Q}^a{\cal Q}^a \rangle$, of ${\cal Q}^a$
vanish when computed with this Gaussian, it follows that
${\cal Q}^a=0$, which physically means that the saturated gluons 
are globally colour neutral.
In fact, since the Wilson lines average to zero over a finite distance
$\Delta x_\perp \simge 1/Q_s(\tau)$, it follows that
colour neutrality is achieved already over a transverse scale of the order
of the saturation length (this can be explicitly verified by using eq.~(\ref{lamRPA})):
\be
\int d{\rm y} 
%_{\tau_s(\kk)}^\tau d{\rm y} 
\int_{\Delta S_\perp} \!{d^2x_\perp} \,\rho_{\rm y}^{a}(x_\perp)
\,=\,0,\ee
where $\Delta S_\perp$ is, e.g., a disk of radius $R > 1/Q_s(\tau)$.

When $\kk\to 0$, the spectrum (\ref{lamRPA}) vanishes
sufficiently fast to ensure the infrared finiteness of the
dipole scattering amplitude (\ref{Stau-MOM}), and, more generally,
of all the quantities which were, at most, linearly infrared divergent
in the MV model. Thus, we expect that all the observables computed
within the CGC effective theory come out infrared finite, and thus are
insensitive
to the non-perturbative physics at momenta $\kk\simle \Lambda_{QCD}$.
In fact, because of saturation, the spectrum starts to soften
already at momenta $\kk > Q_s$, where the linear, BFKL, evolution
applies. This happens in the window for ``geometric scaling'' 
\cite{SCALING}, to be discussed in Sect. 6.

\subsection{Brief summary of the quantum evolution}

To summarize, if we compare to the initial condition of the MV
type --- namely, the local Gaussian (\ref{WMV}), which describes a
system of independent colour sources ---, the general effect of the quantum
evolution is to introduce {\it correlations} in the transverse and
longitudinal directions (i.e., non--localities in $r_\perp$ and y), 
and {\it non--linearities}
in $\rho$ (i.e., higher $n$--point correlations
among the colour sources). However, the role and the relative
importance of the various non--local effects, and of the
higher $n$--point functions, may differ from one
kinematical regime to another.

At transverse momenta far above  $Q_s$, the longitudinal correlations
are irrelevant (they are always integrated out, as in 
eqs.~(\ref{rhoT}) and (\ref{vlinear})), and the various 
$n$--point correlations evolve independently from each other.
In particular, the $2$--point function evolves according to the
BFKL equation (\ref{BFKL}), and so does the dipole--hadron
scattering amplitude.

In the opposite limit at $\kk\simle Q_s$, the RGE becomes linear
(cf. eq.~(\ref{RGE-RPA})), and the weight function is strictly Gaussian,
and therefore local in y. The only correlations which persist
in this regime are the transverse correlations reflecting the
{colour neutrality} of the saturated gluons. 
These are sufficient to ensure that gauge--invariant observables 
come out infrared finite when computed in the effective theory.

In the intermediate regime at $\kk\sim Q_s$, the evolution is
fully non--linear --- correlation functions of all orders get mixed in 
the evolution equations ---, and therefore very non--local in y.
Clearly, there is no hope to describe the complex physics in this
transition regime via a Gaussian approximation. However, we expect
this physics not to be essential for physical quantities for which 
the external resolution scale is either well above, 
or well below, the saturation scale. As we shall verify on the
example of the dipole--hadron scattering amplitude in Sect. 7.2,
such quantities receive most of their contributions from either the 
hard modes with $\kk\gg Q_s$, or the saturated
gluons with $\kk\ll Q_s$.

To easily compute such observables, it would be advantageous to dispose
of a Gaussian weight function which interpolates smoothly between
the known behaviours %of the $2$--point function 
at high and, respectively, low transverse momenta. For instance, the 
complete solution $\lambda_\tau(x_{\perp}, y_\perp)$
to the evolution equation in the MFA, eq.~(\ref{lambdaeq}),
would provide such an interpolation, since this is necessarily
a smooth function and has the right limiting behaviours.
But eq.~(\ref{lambdaeq}) is a complicated,
non--linear, integral equation, whose general solution
can be obtained, at most, numerically. For practical
calculations, it would be preferable to have a simple analytic
interpolation, which satisfies the BFKL  equation (\ref{BFKL})
at  high $\kk$, and reduces to eq.~(\ref{lamRPA}) at low $\kk$,
with the transition between the two regimes taking place
at  $\kk\sim Q_s$. Such  an interpolation will be constructed in 
the next section.

\section{The interpolation}
\setcounter{equation}{0}

So far, we have found the following limiting behaviours for the charge--charge 
correlator (\ref{rhoHM}) at high and, respectively, low transverse momenta:
\be\label{lam-sum}
\bar\lambda_\tau(k_{\perp})\,\approx
\left\{ \begin{array} {c@{\qquad\rm for\qquad}l}
\lambda_\tau(k_{\perp}), & k_{\perp}\gg Q_s(\tau) \\ 
\frac{1}{\pi}\,
k_{\perp}^2, &  k_{\perp}\ll Q_s(\tau)
\end{array}
\right..
\ee
Here, and from now on, we reserve the notation $\lambda_\tau(k_{\perp})$
for the solution to the BFKL equation,
and use $\bar\lambda_\tau(k_{\perp})$ to denote the kernel of
the Gaussian weight function (\ref{Wgauss}). 

Clearly, the following
function provides a smooth interpolation between the limiting behaviours
shown above:
\be\label{barlam}
\bar\lambda_\tau(k_{\perp})\,\equiv\,\frac{
k_{\perp}^2 \lambda_\tau(k_{\perp})} {k_{\perp}^2+\pi 
\lambda_\tau(k_{\perp})}\,.\ee
There is, however, one more constraint that eq.~(\ref{barlam}) must
satisfy in order to be a physically acceptable interpolation:
the change from the high--momentum regime to the low--momentum one
must happen at the saturation scale. % i.e., for $k_{\perp}\sim Q_s(\tau)$.
For the function in eq.~(\ref{barlam}), this condition is satisfied
provided the two terms in the denominator become of similar magnitude
when $k_{\perp}= Q_s(\tau)$ :
\be\label{QSLAM}
\frac{1}{\pi}\,Q_s^2(\tau)\,\simeq\,\lambda_\tau(k_{\perp}=Q_s(\tau))\,.\ee
Note that this is precisely the condition that the limiting behaviours
shown in eq.~(\ref{lam-sum})
match quasi--continuously with each other at the saturation scale.
Clearly, whether this condition is fulfilled or not, depends 
upon our definition of the saturation scale.
Usually, $Q_s(\tau)$ is defined through a ``saturation criterion'',
imposed on the gluon ``packing factor'' in the 
transverse plane \cite{GLR,MQ85,BM87,AM2},  or, equivalently, 
on the dipole--hadron scattering amplitude 
\cite{AM0,LT99,SAT,SCALING,MT02,FB,DT02}.
For instance, in terms of ${\cal N}_\tau(r_\perp)$, 
the saturation criterion reads ($\kappa$ is a number of order unity)
\be\label{Nsat}
{\cal N}_\tau(r_\perp)\,=\,\kappa
\qquad {\rm for}\qquad r_\perp^2\,=\,1/Q_s^2(\tau)\,,\ee
which is the same as the condition that the dipole
of transverse size equal to the saturation length is completely absorbed
(``blackness'').

We shall argue below that, in fact, eq.~(\ref{QSLAM}) is
just another definition of the saturation scale, equivalent to that
in eq.~(\ref{Nsat}). In other terms, the condition (\ref{QSLAM}) is
satisfied at the right saturation scale {\it by definition}\,,
so eq.~(\ref{barlam}) is a physically acceptable interpolation indeed.
Of course, such an interpolation is not unique, but at the present
level of accuracy we cannot distinguish between different interpolations
which are all physically acceptable according to the previous criteria. 
In what follows, we shall adopt the interpolation (\ref{barlam})
because of its simplicity.

At a mathematical level, the equivalence between the two formulations of
the saturation condition --- in terms of the scattering amplitude, 
eq.~(\ref{Nsat}),
or in terms of the charge-charge correlator, eq.~(\ref{QSLAM}) ---
is almost obvious: Both ${\cal N}_\tau(r_{\perp})$ and 
$\lambda_\tau (\kk)$ satisfy the BFKL equation and obey the same
saturation criterion, namely, the relevant ``scaling function'' 
%(see below)
(${\cal N}_\tau(r_{\perp})$ in one case, and
$\lambda_\tau (\kk)/\kk^2$ in the other)
is constant on the saturation line $\kk = Q_s(\tau)$.
Thus, it is a priori clear that eq.~(\ref{QSLAM}) will provide
the correct estimate for $Q_s(\tau)$. Still, it is instructive to
briefly develop the consequences of this equation here, 
in order not only to recover the known result for 
the saturation scale \cite{AM2,SCALING}, but
also to establish a simple scaling approximation for
$\lambda_\tau(k_{\perp})$, which will greatly simplify 
the use of eq.~(\ref{barlam}) in practice.

For convenience, in what follows we shall mostly consider the solution to the BFKL 
equation (\ref{BFKL}) for the case of a {\it fixed} coupling $\alpha_s$,
and only briefly indicate the changes which occur when a running coupling 
$\alpha_s(Q_s^2)$ is used instead. Also, we shall need 
only the approximate form of the solution, which is obtained via a
saddle point approximation in the Mellin transform of $\lambda_\tau(k_{\perp})$
(see, e.g., \cite{SCALING} for details):
\be\label{lamBFKL}
\lambda_\tau(k_{\perp})% \,\equiv \,{\partial \mu_\tau\over \partial \tau}
\,\simeq\,f(\tau)
%\Phi\Big(\tau,\ln (k_{\perp}^2/Q_0^2)\Big)\,
\sqrt{{Q_0^2}\,{k_{\perp}^2}}\,
\, \exp\Bigg\{\omega\bar\alpha_s\tau-
\frac{1}{2\beta\bar\alpha_s\tau}\left(\ln \frac{k_{\perp}^2}{Q_0^2}\right)^2
\!\Bigg\}\,.\,\,\ee
In this equation,
$f(\tau)$ is a slowly varying function
that will be ignored in what follows (since its contribution
to $\ln(\lambda_\tau/k_{\perp}^2)$  is comparatively small), 
$Q_0$ is a scale  of order $\Lambda_{QCD}$ %which is
introduced by the initial conditions, 
$\bar\alpha_s\equiv\alpha_s N_c/\pi$, 
and $\omega$ and $\beta$ are pure numbers: 
$\omega\!=\!4\ln 2\approx 2.77$, $\beta\!=\!28\zeta(3)\approx 33.67$.

Eq.~(\ref{lamBFKL}) is strictly correct for $k_{\perp}\gg Q_s(\tau)$,
where the colour charge density is low ($\lambda_\tau/k_{\perp}^2\ll 1$),
and BFKL is a right approximation.
But this expression can be extrapolated down to $k_{\perp}\sim Q_s(\tau)$
(since BFKL remains marginally valid down to the saturation line),
and then inserted in the saturation condition (\ref{QSLAM}),
to obtain an equation for $Q_s(\tau)$ :
%, the saturation condition (\ref{QSLAM}) becomes:
\be\sqrt{\frac{Q_0^2}{Q_s^2}}\,
\, \exp\Bigg\{\omega\bar\alpha_s\tau-
\frac{1}{2\beta\bar\alpha_s\tau}\left(\ln \frac{Q_s^2}{Q_0^2}\right)^2
\Bigg\}\,\simeq\,1.\ee
This has the following solution :
\be\label{Qs0}
Q_s^2(\tau)\,=\,Q_0^2 \,
{\rm e}^{c\bar\alpha_s \tau}\,,\qquad c\,=\,%\frac{1}{2}
\big[-{\beta}+\sqrt{\beta(\beta+8\omega)}\,\big]/2\,=\,4.84...\ee
which is the correct result \cite{AM2,SCALING}, as anticipated.
Note that the factor in front of the exponential in eq.~(\ref{Qs0})
is not under control in the present approximations. 
A more refined treatment \cite{MT02,DT02}
shows that this prefactor has actually a weak dependence upon $\tau$. 

It is further interesting to study the behaviour of 
$\lambda_\tau(k_{\perp})$  for momenta close to the
saturation scale, where a simple approximation to 
eq.~(\ref{lamBFKL}) can be obtained. To this aim, let us
replace $Q_0^2$ by $Q_s^2(\tau)$ 
as the reference scale in eq.~(\ref{lamBFKL}), by writing:
\be\label{LQS}
\ln \frac{k_{\perp}^2}{Q_0^2}\, = \,
\ln \frac{k_{\perp}^2} {Q^2_s(\tau)} + c \bar\alpha_s\tau\,.\ee
A simple calculation yields then:
\be\label{BFKL_QS}
\lambda_\tau(k_{\perp}) \,=   \,\frac{1}{\pi}\,  \kk^2\,
\exp\Bigg\{- \gamma\ln \frac{k_{\perp}^2}{Q_s^2(\tau)}
-\frac{1}{2\beta\bar\alpha_s\tau}
\left(\ln \frac{k_{\perp}^2}{Q_s^2(\tau)}\right)^2
\Bigg\},
\ee
where $\gamma\equiv 1/2+c/\beta\approx 0.64$, and the 
(a priori, uncontrolled) prefactor has been chosen in such a way
that the matching condition (\ref{QSLAM}) is fulfilled
exactly\footnote{This is, of course, just a matter of convenience; 
an approximate matching, as shown in eq.~(\ref{QSLAM}), 
would be enough for the present purposes.}.

The last equation suggests a remarkable simplification:
Assume that $k_{\perp}$  is sufficiently close to 
$Q_s(\tau)$ (although still above it) for
$\ln(k_{\perp}^2/Q_s^2(\tau))\ll \bar\alpha_s\tau$.
Then,  the second term in the exponent can be neglected
compared to the first one, 
and  eq.~(\ref{BFKL_QS}) reduces to %can be approximated as
\be\label{lamscaling}
\lambda_\tau(\kk)\, \simeq\, \frac{1}{\pi}\,  \kk^2
\left(\frac{Q_s^2(\tau)}{\kk^2}\right)^\gamma,
\ee
which shows {\it geometric scaling} : %in this approximation,
the dimensionless
function $\lambda_\tau (\kk)/\kk^2$  depends upon its two variables
$\kk^2$  and $\tau$ only via the combination $\kk^2/Q_s^2(\tau)$.
Eq.~(\ref{lamscaling}) is a correct approximation for
momenta $\kk^2$ within the following ``scaling window'' \cite{SCALING}:
\be\label{swindow}
Q_s^2(\tau)\,\,\ll\,\, \kk^2\,\,\ll\,\,\frac{Q_s^4(\tau)}{Q^2_0}\,.\ee

The scaling form (\ref{lamscaling}) is similar to the one derived
in Ref. \cite{SCALING} for the dipole-hadron scattering amplitude,
which reads (for $1/\rr^2$ within the scaling window (\ref{swindow})):
\be\label{Nscaling}
{\cal N}_\tau(r_{\perp})\,\approx\,\kappa \,
\left(r^2_{\perp}Q_s^2(\tau)\right)^\gamma\ee
with the same exponent $\gamma$ as in eq.~(\ref{lamscaling}). 
This exponent (or, rather, the difference $1-\gamma$)
is sometimes referred to as an ``anomalous dimension'' \cite{GLR,BL92,LT99}.
In fact, as we shall check in Sect. 7.2,
the similitude between eqs.~(\ref{lamscaling}) and (\ref{Nscaling}) 
is consistent with the linear
relation (\ref{Nlinear}) between these two quantities,
valid for small $\rr$.

At this point, we should recall that geometric scaling has been
first identified in the experimental data, as a regularity of the HERA data
for $\gamma^*$--proton deep inelastic scattering \cite{geometric}.
More recently, this notion has found phenomenological applications
also to electron--nucleus DIS \cite{eAscaling},
and to heavy ion collisions \cite{KLM02}.

\begin{figure}[htb]
  \centerline{
  \epsfsize=0.7\textwidth
  \epsfbox{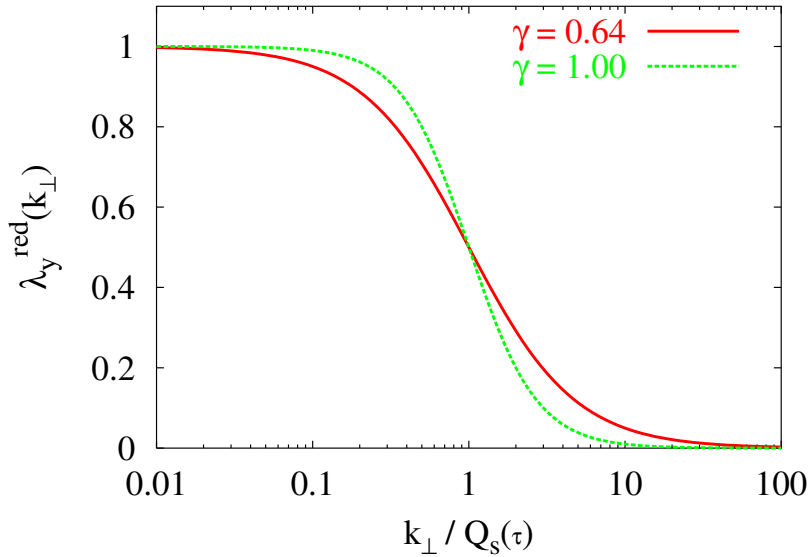}
  }
 \caption[]{Momentum dependence of $\lambda^{\rm red}_{\rm y}(\kk)=
   (\pi/\kk^2)\bar\lambda_{\rm y}(\kk)$, plotted as a function of $\kk/Q_s(\tau)$
for two values of $\gamma$. 
  % Plotted as a function of $\kk/Q_s(\tau)$. 
  Solid (red) line: $\gamma=0.64$, and dashed (green) line 
   $\gamma=1$.}
\label{F1}
\end{figure}

\begin{figure}[htb]
  \centerline{
  \epsfsize=0.7\textwidth
  \epsfbox{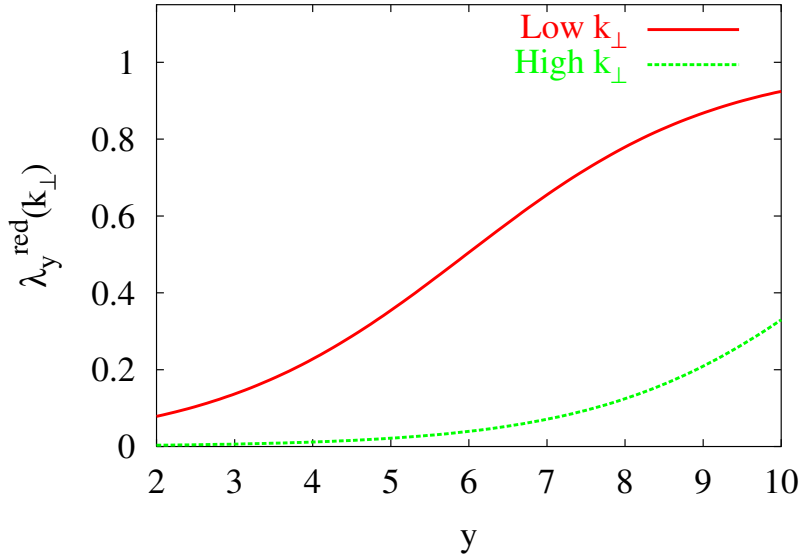}
  }

\caption[]{Longitudinal structure of $\lambda^{\rm red}_{\rm y}(\kk)=
  (\pi/\kk^2)\bar\lambda_{\rm y}(\kk)$. This is plotted as a function of y
 for two transverse momenta:
 $\kk^{\rm high}=3 \, Q_s^2(\tau)$ (green line, dashed), 
 and $\kk^{\rm low}=0.02\, Q_s^2(\tau)$ (red line, solid).
 We have used $\gamma=0.64$, $c=4.84$,  $\bar\alpha_s=0.2$, and $\tau=10$. 
 The corresponding separation rapidities are $\tau_s(\kk^{\rm high})\simeq 11$ 
 and $\tau_s(\kk^{\rm low})\simeq 5$.
   }
\label{F2}
\end{figure}

The previous considerations show that the following scaling Ansatz for the
interpolating kernel (\ref{barlam}) should be a good approximation
for all momenta $\kk\simle Q_s^2(\tau)/\Lambda_{QCD}$:
\be\label{bar2}
\bar\lambda_{\rm y}(k_{\perp})\,=\,\theta(\tau-{\rm y})\,
\frac{k_{\perp}^2}{\pi}\,\frac{
\left(\frac{Q_s^2({\rm y})}{\kk^2}\right)^\gamma}{1 +
\left(\frac{Q_s^2({\rm y})}{\kk^2}\right)^\gamma}
\,.\ee
We have reintroduced here the $\theta$--function to emphasize that
the charge--charge correlator vanishes for space--time rapidities
${\rm y} > \tau$. The quality
of this interpolation will be tested in the next section,
where we shall use eq.~(\ref{bar2}) to compute  
the dipole--hadron scattering amplitude and the unintegrated
gluon distribution, and then we shall compare the results to
known general properties. 

In such applications, it will be convenient to
treat the ``anomalous dimension'' $\gamma$, and also 
the exponent $c$ in eq.~(\ref{Qs0}) for the saturation scale,
as free parameters. For instance, 
eq.~(\ref{bar2}) with $\gamma=1$ provides a simple
generalization of the MV model (in the sense that the transverse
correlations die out at high $\kk$), but which
includes the correct infrared behaviour due to  saturation
(thus avoiding the infrared problem of the original
MV model), and also some of the quantum evolution, via the rapidity
dependence of the saturation scale.

Moreover, eq.~(\ref{bar2}) with $\gamma=1$ can be used as a rough
approximation for the general kernel (\ref{barlam}) in the DLA 
regime\footnote{Note that this is the regime where the dipole--hadron
scattering shows ``colour transparency'' \cite{CTRAN}, i.e., the
corresponding scattering amplitude behaves like 
${\cal N}_\tau(r_\perp)\sim \alpha_s \rr^2\,x G(x,1/\rr^2)/\pi R^2$, where
$x G(x,Q^2)$ is the gluon distribution to be discussed in Sect. 7.1.}
at very high transverse momenta $\kk\gg Q_s^2(\tau)/\Lambda_{QCD}$.
Indeed, in that regime, the solution to the BFKL equation can be replaced
by its DLA approximation, which has only a weak dependence upon $\kk$.
This suggests that, in applications to the phenomenology, one can
always use the simplified kernel (\ref{bar2}), but let the 
anomalous dimension $\gamma$ be weakly dependent upon $\kk$,
in such a way that $\gamma \approx \gamma_{\rm BFKL}\simeq 0.64$ 
within the scaling window
(\ref{swindow}), and $\gamma \approx 1$ for much larger momenta.
More generally, one can imagine extracting both $\gamma$ and $c$
from fits to the experimental data. In the subsequent applications,
we shall give results for both $\gamma=1$ and $\gamma= 0.64$.
%\approx \gamma_{\rm BFKL}$.

The quantity $\bar\lambda_{\rm y}(k_{\perp})$ is represented in Fig. \ref{F1}
as a function of $k_{\perp}$ for fixed y, and in Fig. \ref{F2} as a function
of y for two values of $k_{\perp}$ (one above $Q_s(\tau)$, the other one
below it). For convenience, we have plotted the rescaled function
$\lambda_{\rm y}^{\rm red}(k_{\perp})\equiv 
(\pi/\kk^2)\bar\lambda_{\rm y}(k_{\perp})$.
Note that the y dependence of $\lambda_{\rm y}(k_{\perp})$ represents the
longitudinal profile of the colour charge distribution in the hadron,
for modes of given $\kk$.
(Recall that $\lambda_{\rm y}(k_{\perp})$  is the average colour charge 
squared per unit space--time rapidity per unit transverse area, for a given
$\kk$.) If $\kk > Q_s(\tau)$, the colour charge density increases exponentially
with y all the way up to the edge of the hadron (located at ${\rm y}=\tau$).
If $\kk < Q_s(\tau)$, there is an exponential increase up to the intermediate
rapidity ${\rm y}=\tau_s(\kk)$, where the gluon modes with momentum $\kk$
start to be saturated (cf. eq.~(\ref{taus})); then, for rapidities
$\tau_s(\kk)<{\rm y}<\tau$, the charge density remains nearly constant.
This behaviour is manifest on Fig. \ref{F2}.

To conclude this section, let us indicate how the previous
results change when, instead of a fixed coupling $\alpha_s$,
one rather uses a  coupling which is running with the saturation
momentum: $\alpha_s\longrightarrow \alpha_s(Q_s^2)$, with
$\alpha_s(Q^2)\equiv b_0/\ln(Q^2/\Lambda_{\rm QCD}^2)$.
As shown in Refs. \cite{SCALING,MT02},
the only change refers to the functional form of
$Q_s^2(\tau)$, which now becomes $Q_s^2(\tau)=\Lambda_{\rm QCD}^2 {\rm e}^{\sqrt
{2 b_0 c(\tau+\tau_0)}}$, where $c$ is the same number as in eq.~(\ref{Qs0}),
and $\tau_0$ is a constant. All the other results related to scaling 
--- the value of the anomalous dimension $\gamma$ and the momentum
range (\ref{swindow}) in which the scaling holds --- remain unchanged,
except for the expression of the saturation scale which enters these
results. In the applications to be considered in the
next section, we shall restrict ourselves to the case of a fixed coupling.

\section{Some applications}
\setcounter{equation}{0}

In this section, we shall consider a few applications of the 
Gaussian effective theory with kernel (\ref{bar2}). First, we shall
discuss the unintegrated gluon distribution, for which we shall
derive a simple analytic formula which interpolates between saturation
at low momenta and BFKL evolution at high momenta, and which can be
easily used in further applications. In this context, we shall briefly recall
how saturation arises in the effective theory for the CGC \cite{Cargese}, 
and emphasize
the difference in this respect between the classical MV model \cite{JKMW97,KM98}
and the full theory including quantum evolution \cite{SAT}.
Then, we shall use eq.~(\ref{Stau-MOM}) to compute the
dipole--hadron scattering amplitude and study its various limits.
The resulting expression has the right scaling behaviour at short
distances, and the correct approach towards ``blackness'' at large distances,
and can be easily implemented in phenomenological studies 
of deep inelastic scattering.

\subsection{The unintegrated gluon distribution}

From eqs.~(\ref{bar2}) and (\ref{Qs0}), it is easy to derive the following
approximation for
the quantity $\mu_\tau(\kk)$ which characterizes the distribution
of the two-dimensional colour charge density 
$\rho^a(x_{\perp})\equiv \int\! d{\rm y}\,\rho_{\rm y}^a (x_{\perp})$
(cf. eqs.~(\ref{mutau})--(\ref{mutauMFA})) :
\be\label{mubar}
\bar\mu_\tau(k_\perp)\,=\, \int_{-\infty}^\tau  d{\rm y}\,
\bar\lambda_{\rm y}(k_\perp)\,=\,\frac{k_{\perp}^2}
{\pi\gamma c\bar\alpha_s}\,
\ln\left(1 \,+\,\left(\frac{Q_s^2(\tau)}{\kk^2}\right)^\gamma\right)\,.\ee
This quantity is interesting for, at least, two reasons:

First, it determines the probability distribution for the colour 
charge density in the transverse plane (i.e., the three--dimensional
 distribution integrated over y) :
\be\label{Zgauss}
{Z}_{\tau}[\rho^a(x_{\perp})]
\,=\,{\cal N}_\tau\,
{\rm exp}\!\left\{-\,{1 \over 2}
\int_{x_\perp,y_\perp}\!\frac{\rho^a(x_{\perp})
\rho^a(y_{\perp})}{\bar\mu_\tau(x_{\perp},y_\perp)}
\right\}\,\,.\ee
This is the relevant weight function for applications which
do not consider explicitly the longitudinal structure of the colour source
 (like the numerical simulations in Refs. \cite{KV}).
 
Second, eq.~(\ref{mubar}) provides a good approximation for the 
unintegrated gluon distribution in a wide range of transverse momenta.
More precisely,  $\bar\mu_\tau(\kk)/\kk^2$ coincides with the gluon 
distribution computed in the CGC effective theory
at both high and low momenta compared to $Q_s(\tau)$, and it provides
a smooth interpolation between these regimes.

To show this, we start with the definition of the gluon distribution 
as the Fock-space gluon density in an appropriate gauge.
Consider the canonical quantization
of the Yang--Mills field theory in the light cone gauge $A^+_a=0$ \cite{KS70}. 
The gluon  distribution $x G(x,Q^2)$ (= the number of gluons of
transverse size $\Delta x_\perp \sim 1/Q$ per unit rapidity) is then
obtained as \cite{AMlectures,Cargese} (with $k^+=xP^+$, $\tau=\ln(1/x)$) :
\be\labe{GDFdef}
x G(x,Q^2)&=&\int {d^2k_\perp}\,\Theta(Q^2-k_\perp^2)\,\,
k^+\frac{d^3N}{dk^+d^2k_\perp}\,.\ee
The quantity $d^3N/d^3 k$, which in eq.~(\ref{GDFdef}) plays the role
of the unintegrated gluon distribution, is the Fock space gluon density, 
%i.e., the number of gluons per unit of volume in momentum space, 
which is defined
in terms of the gluon creation and annihilation operators in the standard way.
This can be related to the following %  equal-time
 2--point function of the LC--gauge  ``electric field'' $F^{+i}_a$ :
\be\label{TPS}
%{\cal N}_\tau(k_\perp)\,=\,\frac{1}{\pi R^2}\,
\frac{d^3 N}{d\tau d^2k_\perp}\,=\,\frac{1}{4\pi^3}\,
\langle F^{i+}_a(x^+,k^+,{ k}_\perp)F^{i+}_a(x^+,-k^+,{- k}_\perp) \rangle_\tau,\ee
which is in turn recognized as the LC--gauge expression of a 
gauge--invariant correlator, which in other gauges would involve also Wilson
lines in the longitudinal direction \cite{PI,Cargese}.

In fact, in the effective theory for the CGC, one exploits
this gauge freedom to rewrite the correlation function in eq.~(\ref{TPS})
in terms of the gauge fields in the {\it covariant} gauge,
for which we know the weight function. Namely, we use
eq.~(\ref{Fi+}) to write:
\be\label{UU} 
\langle F_a^{i+}(\vec{x})F_a^{i+}(\vec{y})\rangle_\tau &=& 
\langle (U^\dag_{ab}\del^i\alpha^b)_{\vec{x}}\,
(U^\dag_{ac}\del^i\alpha^c)_{\vec{y}}\rangle_\tau\,\ee
with $\vec{x}=(x^-,x_\perp)$, and $U^\dag(\vec{x})$ and
$\alpha_a(\vec{x})$ given
by eqs.~(\ref{UTA}) and (\ref{alpha}), respectively. 
The  Wilson lines in the r.h.s. of eq.~(\ref{UU}) encode all
the non-linearities coming from the classical field equations (\ref{cleq}).

Within the Gaussian effective theory with weight function (\ref{Wgauss}),
this correlation function can be explicitly computed, with the following result 
\cite{SAT,Cargese} :
\be\label{Fock}
\frac{1}{\pi R^2}\,
\frac{d^3 N}{d\tau d^2k_\perp}\,=\,
\frac{N_c^2-1}{4\pi^3}\int^\tau_{-\infty} d{\rm y} \int d^2 r_\perp \,
{\rm e}^{i\kk\cdot r_\perp } \,S_{\rm y}(r_\perp)
 \left(-\nabla_\perp^2 \gamma_{\rm y}(r_\perp ) 
\right),\ee
where $\gamma_{\rm y}(r_\perp )$ is the 2--point function
of the field $\alpha$, eq.~(\ref{gamlam}), 
and $S_{\rm y}(r_\perp)$ is the 2--point function
of the Wilson lines in the adjoint representation, i.e.,
eq.~(\ref{Stau-MOM}) with $C_R= N_c$.

To simplify notations, it is useful to define:
\be\label{phidef}
\varphi_\tau(\kk)\,\equiv\,\frac{4\pi^3}{N_c^2-1}\,
\frac{1}{\pi R^2}\,\frac{d^3 N}{d\tau d^2k_\perp}\,,\ee
a quantity that we shall refer to as the
``unintegrated gluon distribution'' in what follows. Up to the factor
$4\pi^3$, this is the number of gluons of each colour per unit rapidity
per unit of transverse phase--space.

Clearly, even with the interpolation (\ref{bar2}) for the kernel
of the Gaussian, which implies 
\be\label{gammabar}
\bar\gamma_{\rm y}(k_\perp)\,=\,
\frac {1}{\pi k_{\perp}^2}\,\frac{
\left(\frac{Q_s^2({\rm y})}{\kk^2}\right)^\gamma}{1 +
\left(\frac{Q_s^2({\rm y})}{\kk^2}\right)^\gamma}\,,\ee
(cf. eq.~(\ref{gammak}))
it is still not easy to perform the remaining integrations
in eq.~(\ref{Fock}) analytically. But simpler formulae
can be obtained in the limiting regimes where $k_\perp$ 
is either well above, or well below, the saturation scale.

Consider the high--$k_\perp$ regime first. 
Then, the integral in eq.~(\ref{Fock}) is dominated by small
$r_\perp\ll 1/Q_s(\tau)$, so that $S_{\rm y}(r_\perp)\approx 1$, 
and the r.h.s. of eq.~(\ref{Fock}) reduces to 
\be\label{intyk}
\int^\tau_{-\infty} d{\rm y} \int d^2 r_\perp \,
{\rm e}^{i\kk\cdot r_\perp } \,
 \left(-\nabla_\perp^2 \gamma_{\rm y}(r_\perp ) 
\right)\,=\,\int^\tau_{-\infty} d{\rm y}\, \,\kk^2\gamma_{\rm y}(k_\perp )
\,=\,\frac{\mu_\tau(k_\perp)} {\kk^2}\,,\ee
where we have also used eqs.~(\ref{gammak}) and (\ref{mutauMFA}). This gives:
\be\label{Fock-high}
\varphi_\tau(\kk)
\,\simeq\,\,\frac{\mu_\tau(k_\perp)} {\kk^2}\,
\qquad  {\rm for} \qquad k_\perp\gg Q_s(\tau),
\ee
with $\mu_\tau(k_\perp)$ obtained as the solution to the BFKL
equation (\ref{BFKL}). Eq.~(\ref{Fock-high}) is the standard
BFKL approximation for  the unintegrated gluon distribution.

\vspace*{.2cm}
The low--$k_\perp$ case is more subtle, and we shall construct
our argument in several steps:

(I) We shall see eventually that, when  $k_\perp\ll Q_s(\tau)$, the dominant
contribution to eq.~(\ref{Fock})
comes from space-time rapidities in the range
$\tau_s(\kk)<{\rm y}<\tau$, with $\tau_s(\kk)$ defined in eq.~(\ref{taus}).
In order to check this, we first evaluate the contribution of the
lower rapidities, ${\rm y}<\tau_s(\kk)$.  In this range, $Q_s({\rm y})
\ll \kk$, and the exponential factor ${\rm e}^{i\kk\cdot r_\perp }$ 
selects distances $r_\perp\simle 1/k_\perp \ll 1/Q_s({\rm y})$,
for which $S_{\rm y}(r_\perp)\simeq 1$ to a very good accuracy.
Thus, this low--y contribution is proportional to (cf. eq.~(\ref{intyk}))
\be\label{subdom}
\int^{\tau_s(\kk)}_{-\infty} d{\rm y} \int d^2 r_\perp \,
{\rm e}^{i\kk\cdot r_\perp } \,
 \left(-\nabla_\perp^2 \gamma_{\rm y}(r_\perp ) 
\right)\,=\,\frac{\mu_{\tau}(k_\perp)} {\kk^2}\bigg|_{\tau=\tau_s(\kk)}
\,\sim\,
\frac{1}{\bar\alpha_s}\,,\ee
where the last estimate comes from using eq.~(\ref{mubar}).
The contribution (\ref{subdom}) 
is a priori large, since of order $1/\alpha_s$, but nevertheless
subdominant, as we shall see.

(II) We now consider the remaining contribution, namely:
\be\label{dom}
\int^\tau_{\tau_s(\kk)} d{\rm y} \int d^2 r_\perp \,
{\rm e}^{i\kk\cdot r_\perp } \,S_{\rm y}(r_\perp)
 \left(-\nabla_\perp^2 \gamma_{\rm y}(r_\perp ) 
\right),\ee
where, for a typical y, $\kk\ll Q_s({\rm y})\ll Q_s(\tau)$. Since 
 ${\rm e}^{i\kk\cdot r_\perp }$ selects relatively large distances 
$r_\perp\simge 1/Q_s({\rm y})$, we must consider the behaviour
of the integrand for such distances. In momentum space, we have (cf. 
eqs.~(\ref{gammak}) and (\ref{lamRPA}), or directly from eq.~(\ref{gammabar})) :
\be\label{gammalow}
\kk^2\gamma_{\rm y}(k_\perp )\,\simeq\,1/\pi,\qquad {\rm for}\qquad
k_\perp\ll Q_s({\rm y}),\ee
showing that, in coordinate space, 
%the integrand of eq.~(\ref{Fock}) is a $\delta$--function:
$-\nabla_\perp^2 \gamma_{\rm y}(r_\perp ) = \pi \delta^{(2)}(r_\perp )$.
Since eq.~(\ref{gammalow}) is just a low-momentum approximation, 
this simply means that the function
$-\nabla_\perp^2 \gamma_{\rm y}(r_\perp )$ is localized at small distances
$r_\perp\simle 1/Q_s({\rm y})$, with integrated weight
\be
\int d^2 r_\perp \, \left(-\nabla_\perp^2 \gamma_{\rm y}(r_\perp ) 
\right)\,=\,1/\pi\,.\ee
Thus, in reality, the integration in eq.~(\ref{dom}) is restricted to
values $r_\perp\simle 1/Q_s({\rm y})$, where  we cannot compute
the integrand accurately (since our approximations break down around
$Q_s$), but where $S_{\rm y}(r_\perp)\approx \delta < 1$ is essentially 
a number of order one. That is, we shall approximate:
\be\label{dom1}
\hspace*{-.3cm}
\int^\tau_{\tau_s(\kk)}\! d{\rm y}\! \int\! d^2 r_\perp \,
{\rm e}^{i\kk\cdot r_\perp } \,S_{\rm y}(r_\perp)
 \left(-\nabla_\perp^2 \gamma_{\rm y}(r_\perp ) 
\right)\,\approx\,\frac{\delta}{\pi}\,\int^\tau_{\tau_s(\kk)} d{\rm y}\,=\,
\frac{\delta}{\pi}\,\Big(\tau-\tau_s(k_\perp)\Big),\ee
where the number $\delta < 1$ cannot be computed in the present approximations.
Since, moreover (cf. eqs.~(\ref{Qs0}) and (\ref{taus})):
\be\label{tautaus}
\tau-\tau_s(k_\perp)\,=\,\frac{1}
{ c\bar\alpha_s}\,\ln\frac{Q_s^2(\tau)}{\kk^2}\,,
\ee
where, by assumption, $\ln({Q_s^2(\tau)}/\kk^2)\gg 1$, it is clear
that the contribution (\ref{dom1}) is larger than that in eq.~(\ref{subdom}),
and thus is the dominant contribution, as anticipated.

To conclude, at low momenta $k_\perp\ll Q_s(\tau)$, we have found:
\be\label{Fock-low}
\varphi_\tau(\kk)\,\simeq\,
\delta\,\frac{1}
{\pi c\bar\alpha_s}\,\ln\frac{Q_s^2(\tau)}{\kk^2}\,,
\qquad  {\rm for} \qquad k_\perp\ll Q_s(\tau)\,,
\ee
which, up to the factor $\delta$, is the same as $\mu_\tau(\kk)/\kk^2$,
with $\mu_\tau(\kk)$ given by its low-momentum approximation (\ref{mu-sat}).

Eq.~(\ref{Fock-low})
exhibits {\it gluon saturation}. To see this, compare eqs.~(\ref{Fock-high})
and (\ref{Fock-low}): At high momenta,  $\varphi_\tau(\kk)$ shows an
exponential increase  with $\tau$ (according to the
BFKL equation), % (see  eq.~(\ref{lamBFKL}) for an example), 
and also a rapid increase with decreasing $\kk$ (as a power of $1/\kk$),
which, if extrapolated at low momenta, would generate infrared problems.
But for $\kk$ below $Q_s(\tau)$, $\varphi_\tau(\kk)$ is rather described
by eq.~(\ref{Fock-low}), which shows only a {\it linear} increase with $\tau$,
that is, logarithmic with $1/x$  --- this is {\it small--$x$ saturation} ---,
and also a logarithmic  increase with $1/\kk$ --- this is  {\it 
$\kk$--saturation}. The fact that both types of saturation occur simultaneously
is consistent with the notion of ``geometric scaling'': At saturation,
there is only one intrinsic scale in the problem, the saturation scale,
so $\varphi_\tau(\kk)$ can be only a function of the ratio
${Q_s^2(\tau)}/\kk^2$. Thus, a logarithmic dependence upon $\kk$
entails necessarily a similar behaviour in $\tau\propto \ln Q_s^2(\tau)$.
As we have seen in Sect. 6, the scaling property characteristic
of saturation is preserved
by the BFKL evolution until relatively large values of $\kk$.

At this point, let us also emphasize a crucial difference between the saturation 
mechanisms in the MV model and in the full effective theory, 
with quantum evolution
included. In the MV model, $\kk$--saturation occurs as well
\cite{JKMW97,KM98}, but only as a consequence of the non-linear
effects in the  classical field equations (\ref{cleq}).
That is, this classical saturation relies crucially on the presence 
of the Wilson lines in eq.~(\ref{UU}), 
or of their 2--point function $S_{\rm y}(r_\perp)$ in eq.~(\ref{Fock}).

On the other hand, these Wilson lines played almost no role in the
quantum calculation leading to eq.~(\ref{Fock-low})
(they are just responsible for the overall factor $\delta$).
In this case, the saturation is rather the consequence of 
the long-range correlations among the colour sources, as encoded in
the low--$\kk$  behaviour of the function $\mu_\tau(\kk)$
(cf. eqs.~(\ref{mu-sat}) and (\ref{tautaus})).
These correlations ensure the global neutrality of the
colour sources at the scale $1/Q_s(\tau)$, and this in turn
implies the saturation of the radiated gluons {\it independently} of
the non--linear effects in the classical field dynamics.
The relevant non--linear effects are rather those in the quantum evolution,
i.e., in the RGE (\ref{RGE}), which generate the  scale $Q_s(\tau)$
in the first place, as the scale for colour  neutrality.

\begin{figure}%[htb]
  \centerline{
  \epsfsize=0.7\textwidth
\epsfbox{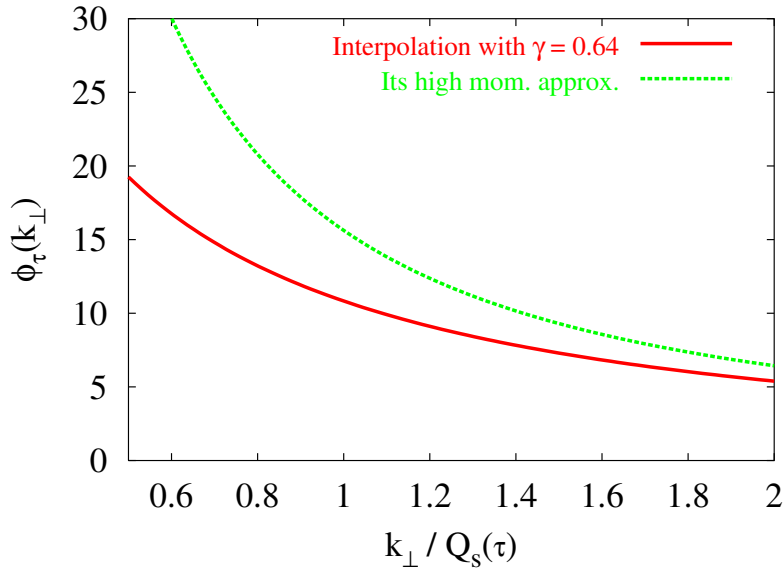}
  }
 \caption[]{$\bar\varphi_\tau(\kk)$ as a function of  $\kk/Q_s(\tau)$ 
  in linear scale (red line, solid), compared with its high momentum 
  approximation $(Q_s^2(\tau)/\kk^2)^{\gamma}$ (green line, dashed). 
  We have used $\bar\alpha_s=0.1$ and $\gamma=0.64$.}
\label{F3}
\end{figure}

\begin{figure}[htb]
  \centerline{
  \epsfsize=0.7\textwidth
\epsfsize=0.7\textwidth
  \epsfbox{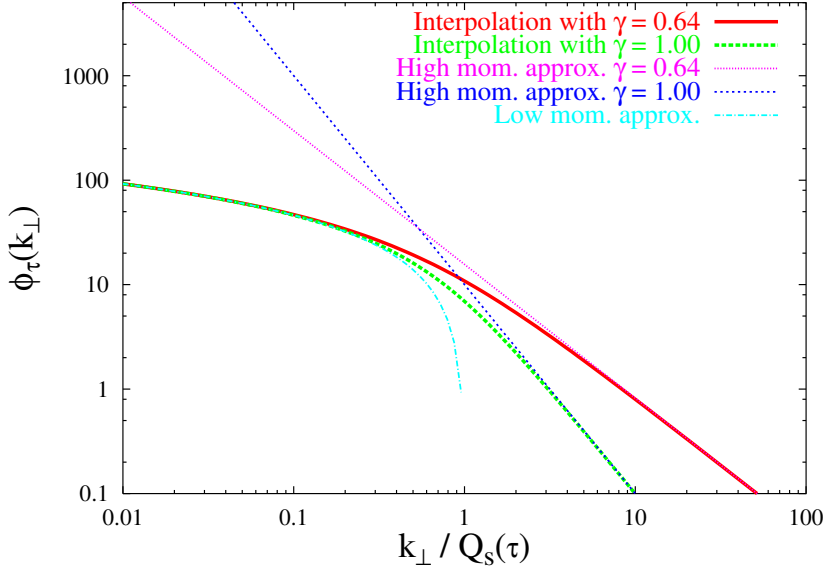}
  }
 \caption[]{$\bar\varphi_\tau(\kk)$ as a function of $\kk/Q_s(\tau)$ 
  in log--log scale for two values of $\gamma$ :
  $\gamma=0.64$ (red line, thick solid), and
  $\gamma=1$ (green line, thick dashed).
  Comparison is made with $1/\kk^{2\gamma},\ \gamma=0.64$ 
  (magenta line, thin dotted), $1/\kk^2$ (blue line, thin dashed),    
  and $(1/\bar\alpha_s)\ln( Q_s^2(\tau)/\kk^2)$ (cyan line, thin dot--dashed).}
\label{F4}
\end{figure}

\begin{figure}%[htb]
  \centerline{
   \epsfsize=0.8\textwidth
  \epsfbox{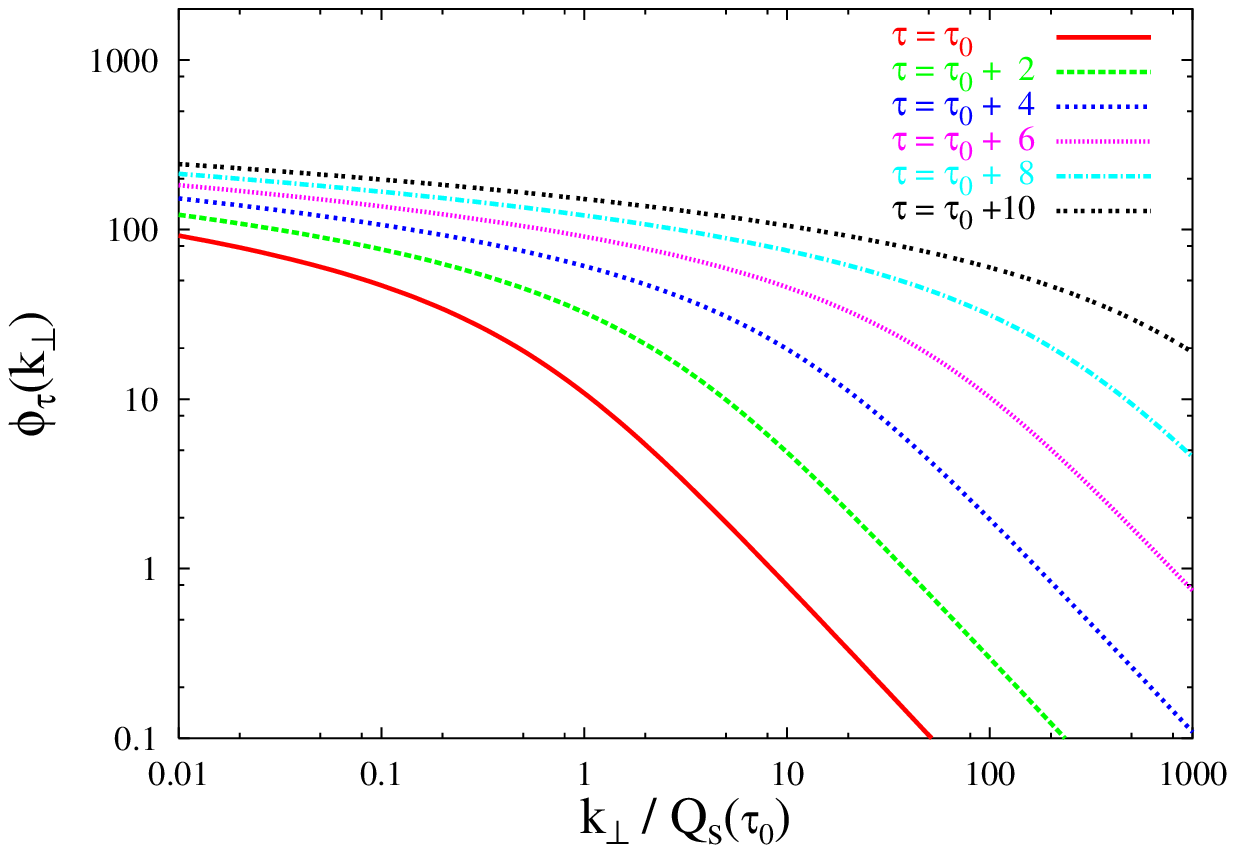}
  }
 \caption[]{Energy dependence of $\bar\varphi_\tau(\kk)$ for
$\gamma=0.64$. We have plotted 
  $\varphi_{\tau}(\kk)$  as a function of $\kk/Q_s(\tau_0)$ (with $\tau_0$
  some value of reference) for six values of $\tau$.
  The lines, from the bottom to the top, correspond successively to 
  $\tau=\tau_0,\, \tau_0 + 2,\, \cdots,\, \tau_0+10.$ The increase 
  with $\tau$ is exponential at high momenta (giving equidistant curves 
  in this log--log plot), but only logarithmic at low  momenta.}
\label{F5}
\end{figure}

These considerations explain why there is no contradiction, 
in this quantum context, between having an expression for the 
unintegrated gluon distribution which is {\it linear} in 
$\mu_\tau(\kk)$ both at high momenta and at low momenta, as we have found
before, and which nevertheless shows saturation. This further implies
that it is enough to dispose of a good approximation for 
the 2--point function $\mu_\tau(\kk)$, namely eq.~(\ref{mubar}), to immediately
deduce an equally good approximation for the unintegrated gluon distribution, 
that is,
\be\label{barphi}
\bar\varphi_\tau(\kk)
\,\equiv\,\,\frac{\bar\mu_\tau(k_\perp)} {\kk^2}\,=\,
\frac{1}{\pi\gamma c\bar\alpha_s}\,
\ln\left(1 \,+\,\left(\frac{Q_s^2(\tau)}{\kk^2}\right)^\gamma\right)\,.\ee
At low momenta, $\kk\ll Q_s(\tau)$, this clearly reproduces the 
behaviour in eq.~(\ref{Fock-low}) (except for the factor $\delta$, which is unknown
anyway, and will be ignored in what follows).
At high momenta,  $\kk\gg Q_s(\tau)$,
\be\label{muhigh}
\bar\varphi_\tau(k_\perp)\,\simeq\,\frac{1}
{\pi\gamma c\bar\alpha_s}\,\left(\frac{Q_s^2(\tau)}{\kk^2}\right)^\gamma,
\ee
which is the geometric scaling version of the BFKL expression in
eq.~(\ref{Fock-high}), and therefore holds,
strictly speaking, only within the scaling window (\ref{swindow}).

According to (\ref{barphi}), the saturation condition can be 
also formulated in terms of $\varphi_\tau\,$: $Q_s(\tau)$ is the scale at
which the  unintegrated gluon distribution becomes of order $1/\alpha_s$:
\be\label{satmu}
\varphi_\tau(\kk)
\,\sim\,\frac{1}{\bar\alpha_s}
\qquad{\rm for}\qquad \kk\sim Q_s(\tau)\,.\ee

The $\kk$--dependence of the function (\ref{barphi}) is 
illustrated in Figs. \ref{F3}, \ref{F4} and \ref{F5}, by using both linear and
logarithmic scales, % (which allows for a wider kinematical region to be shown), 
so that the various regimes alluded to before become manifest.
In particular, in Fig. \ref{F5}, we have indicated also the $\tau$--dependence
of the result, by plotting  $\bar\varphi_\tau(k_\perp)$ as a function
of $\kk$ for several values of $\tau$.

\subsection{The dipole--hadron scattering amplitude}

After using eq.~(\ref{mubar}), the $S$--matrix element for dipole--hadron 
 scattering, eq.~(\ref{Stau-MOM}), becomes [for a gluonic dipole, i.e.,
with $C_R=N_c$ ] :
\BQ\label{barS}
S_\tau(\rr)\,=\,
\exp\left\{ -\frac{4\pi}{\gamma c}
\int \!{d^2k_\perp\over (2\pi)^2}\,\frac{1-
{\rm e}^{ik_\perp\cdot r_\perp}}{k_\perp^2}\,\,
\ln\left[1 \,+\,\left(\frac{Q_s^2(\tau)}{\kk^2}\right)^{\!\gamma}\,\right]
\right\}\,\equiv\,{\rm e}^{-{\Omega}_\tau(\rr)}
.
\EQ

Clearly, for any $\gamma >0$, the integral in the exponent is both
ultraviolet and infrared finite, so, by dimensional analysis,
its result must be a scaling function (i.e., a function of $\rr Q_s(\tau)$).
In what follows, we shall study the asymptotic behaviour of this function
for both $\rr Q_s(\tau)\ll 1$ (small dipole) and 
$\rr Q_s(\tau)\gg 1$ (large dipole). This is interesting for several
reasons: {\it i\,})  It allows us to identify the dominant scattering
mechanisms in these limits.
{\it ii\,}) It illustrates the role of saturation in providing
a smoother infrared behaviour, and thus infrared finite results
(this will be especially clear in the comparison with the MV model
for $\gamma=1$). {\it iii\,}) For a small dipole and $\gamma<1$, we shall
verify explicitly that the anomalous dimension  $\gamma$ is correctly
transmitted from the kernel of the Gaussian to the scattering amplitude
(cf. eq.~(\ref{Nscaling})).  {\it iv\,}) For a large dipole, we shall
recognize the behaviour expected from  previous
studies of the BK equation \cite{LT99,AMlectures}, and of the RGE (\ref{RGE})
\cite{SAT,Cargese}). Also, we shall present 
the results of the numerical evaluation
of eq.~(\ref{barS}) for two values of $\gamma$.

\bigskip
\noindent{\bf (I) Short-range behaviour: $\rr\ll 1/Q_s(\tau)$}

To estimate the integral in eq.~(\ref{barS}), we divide
the integration range into three domains separated by 
the external scales $1/\rr$ and $Q_s(\tau)$. Thus, we need
 to consider the following domains:
(I--a) $1/\rr\ll \kk$, (I--b) $Q_s(\tau)\ll \kk\ll 1/\rr$, and
(I--c) $\kk\ll Q_s(\tau)$.

In domain (I--a), $\kk$ is the largest scale in the problem, so
$\kk\cdot \rr \gg 1$ and  $Q_s^2(\tau)/\kk^2\ll 1$. Then, 
 one can ignore the rapidly oscillating exponential
${\rm e}^{i\kk\cdot\rr}$, and also expand the logarithm as $\ln (1+x)\simeq x$. 
This gives:
\BQ\label{Ia}
\Omega^{\rm (I-a)}\,\simeq \,\frac{1}{\gamma c} \int_{1/\rr^2}^\infty  
\frac{d k^2_\perp}{k_\perp^2}\, \left(\frac{Q_s^2(\tau)}{\kk^2}\right)^\gamma
=\frac{1}{\gamma^2 c}\left(\rr^2Q_s^2(\tau)\right)^\gamma .
\EQ

In domain (I--b), one can expand both the logarithm, as above,
 and the exponential,
and keep only the leading term which survives after the angular 
integral. That is, we can replace
$1-{\rm e}^{i\kk\cdot\rr}\rightarrow \frac14 \rr^2 \kk^2$, and thus obtain:
\BQA\label{Ib}
\hspace*{-0.8cm}
\Omega^{\rm (I-b)}\simeq \,\frac{1}{4\gamma c} \,\rr^2 
\int_{Q_s^2(\tau)}^{1/\rr^2}  
{d k^2_\perp}\left(\frac{Q_s^2(\tau)}{\kk^2}\right)^\gamma
\simeq\left\{
\begin{array} {c@{\quad\rm for\quad}l}
\frac{1}{4c} \rr^2Q_s^2(\tau) \ln (1/\rr^2Q_s^2(\tau))&\gamma=1 \\
a\left(\rr^2Q_s^2(\tau)\right)^\gamma &\gamma<1
\end{array}
\right.
\EQA
with $a =1/[4c\gamma(1-\gamma)]$.
 Note that the case $\gamma=1$ requires a separate treatment.

Lastly, in domain (I--c), we can expand the exponential as above,
and approximate $\ln (1+x)\simeq \ln x$. We obtain:
\BQ
\Omega^{\rm (I-c)}\,\simeq\, \frac{1}{4c}\,\rr^2 \int_0^{Q_s^2(\tau)}
{d k^2_\perp}\, \ln \left(\frac{Q_s^2(\tau)}{\kk^2}\right)
=\frac{1}{4c}\, \rr^2Q_s^2(\tau).
\EQ

Putting the results altogether, it is clear that, when $\gamma=1$, 
the leading contribution comes from domain (I-b), 
since this is enhanced by a large logarithm. This gives:
\BQ\label{Stau_short_gamma_1}
S_\tau(\rr)\bigg|_{\gamma=1}\simeq\, \exp\left\{-\, \frac{1}{4 c}\, \rr^2Q_s^2(\tau) 
\ln \frac{1}{\rr^2Q_s^2(\tau)}\right\} ,
\EQ
which is formally similar to the MV model (compare to eq.~(\ref{SMV1})), with the
crucial difference, however, that the scale in the logarithm  
is now set by the (hard) saturation momentum $Q_s(\tau)$, rather than 
by $\Lambda_{\rm QCD}$. 
This reflects the fact that, due to colour neutrality at saturation,
the quantum effective theory is free of infrared problems.

Note also that the exponent in eq.~(\ref{Stau_short_gamma_1})
vanishes at $\rr= 1/Q_s(\tau)$, which indicates that this
expression is valid only at very short distances, $\rr\ll 1/Q_s(\tau)$,
and thus cannot be used to study the approach towards saturation. For this
latter purpose, a more refined calculation is
necessary, and for $\gamma=1$, this can be even performed analytically. We
 shall return to this  calculation towards the end of the section.

When $\gamma< 1$, on the other hand, it is clear from eqs.~(\ref{Ia})
and (\ref{Ib}) that  domains (I--a) and (I--b) contribute on the same
footing, and give together the leading behaviour. This shows that the integral
in eq.~(\ref{barS}) is now dominated by momenta $\kk\sim 1/\rr$,
as expected for the BFKL dynamics. We thus deduce
that, for $\rr\ll 1/Q_s(\tau)$ (but within the scaling window (\ref{swindow})),
the scattering amplitude has the right scaling form:
\be\label{Oscaling}
{\cal N}_\tau(r_{\perp})\,\approx\,{\Omega}_\tau(\rr)\bigg|_{\gamma<1}\simeq\,
\kappa \,
\left(r^2_{\perp}Q_s^2(\tau)\right)^\gamma\ee
(with $\kappa$ some unknown number) in agreement with eq.~(\ref{Nscaling}).

Note that $\kk$ is the momentum transferred between the incoming dipole,
of transverse size $\rr$, and the colour sources that it scatters off in
the hadronic target. Thus, the previous discussion shows that a small dipole
($\rr\ll 1/Q_s(\tau)$) undergoes mostly {\it hard} scattering
($\kk\gg Q_s(\tau)$), that is, it interacts predominantly
with the colour sources in its vicinity in the impact parameter space.
This agrees with the conclusions in Ref. \cite{FB}.

\bigskip
\noindent{\bf (II) Long distance  behaviour: $\rr \gg 1/Q_s(\tau)$}

The three domains that we have to consider now are:
(II--a) $\kk\gg Q_s(\tau)$, 
(II--b) $Q_s(\tau)\gg \kk\gg 1/\rr$,  and
(II--c) $1/\rr\gg \kk$. Their respective contributions
are easily estimated, with the following results:
\BQA
\Omega^{\rm (II-a)}&\simeq& \frac{1}{\gamma c}
\int^{\infty}_{Q_s^2(\tau)}\frac{d\kk^2}{\kk^2}
\left(\frac{Q_s^2(\tau)}{\kk^2}\right)^\gamma=\frac{1}{\gamma^2 c},\\
\Omega^{\rm (II-b)}&\simeq& \frac{1}{c}\int^{Q_s^2(\tau)}_{1/\rr^2}\frac{d\kk^2}{\kk^2}
\ln \left(\frac{Q_s^2(\tau)}{\kk^2}\right)=\frac{1}{2c}\,\Big( \ln
{\rr^2Q_s^2(\tau)}\Big)^2,\\
\Omega^{\rm (II-c)}&\simeq& \frac{1}{4c} \,\rr^2 
\int^{1/\rr^2}_{0}{d\kk^2}
\ln \left(\frac{Q_s^2(\tau)}{\kk^2}\right)\simeq \frac{1}{4c} \ln \rr^2 Q_s^2(\tau).
\EQA
Clearly, to leading--log accuracy, the  dominant contribution comes 
from domain (II--b), and gives:
\BQ\label{Stau_long}
S_\tau(\rr)\propto 
\exp\left\{-\frac{1}{2c}\Big(\ln {\rr^2Q_s^2(\tau)} \Big)^2\right\}\, ,
\EQ
in agreement with the results in Refs. \cite{LT99,SAT,Cargese,AMlectures}.
In this case, the dipole scattering is dominated by relatively long--range
interactions ($\kk\ll Q_s(\tau)$, with $\kk\gg 1/\rr$ though), that is,
by the interactions with the saturated gluons.

Eq.~(\ref{Stau_long}) describes the approach of the $S$--matrix element 
towards the ``black disk'' limit $S_\tau(\rr)=0$ in the regime
where the dipole is already very large. Thus, when 
this functional form starts to apply, $S_\tau(\rr)$ is 
already significantly smaller than one (i.e., the
proportionality coefficient in eq.~(\ref{Stau_long}) should be rather small).

\bigskip

\begin{figure}[htb]
  \centerline{
  \epsfsize=0.7\textwidth
  \epsfbox{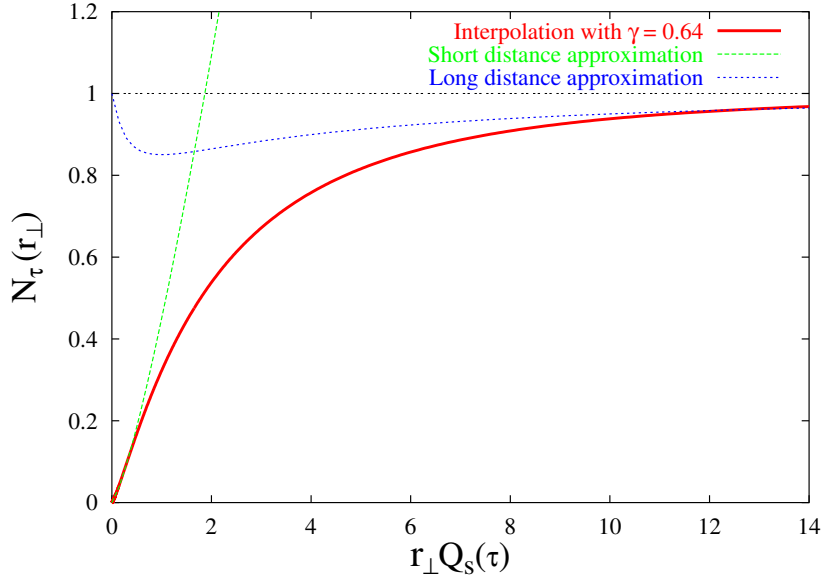}
  }
 \caption[]{%{}\\
      %\hspace*{1.5cm}
      The scattering amplitude $1-S_\tau(r_\perp)$ for $\gamma=0.64$ 
      plotted as a function of $r_\perp Q_s(\tau)$. Red solid line: 
      numerical evaluation of eq.~(\ref{barS});
 green dashed line: short distance 
      approximation, eq.~(\ref{Oscaling});
 blue dotted line:
            long distance approximation, eq.~(\ref{Stau_long}).}
\label{F6}
\end{figure}

\begin{figure}[htb]
  \centerline{
  \epsfsize=0.7\textwidth
  \epsfbox{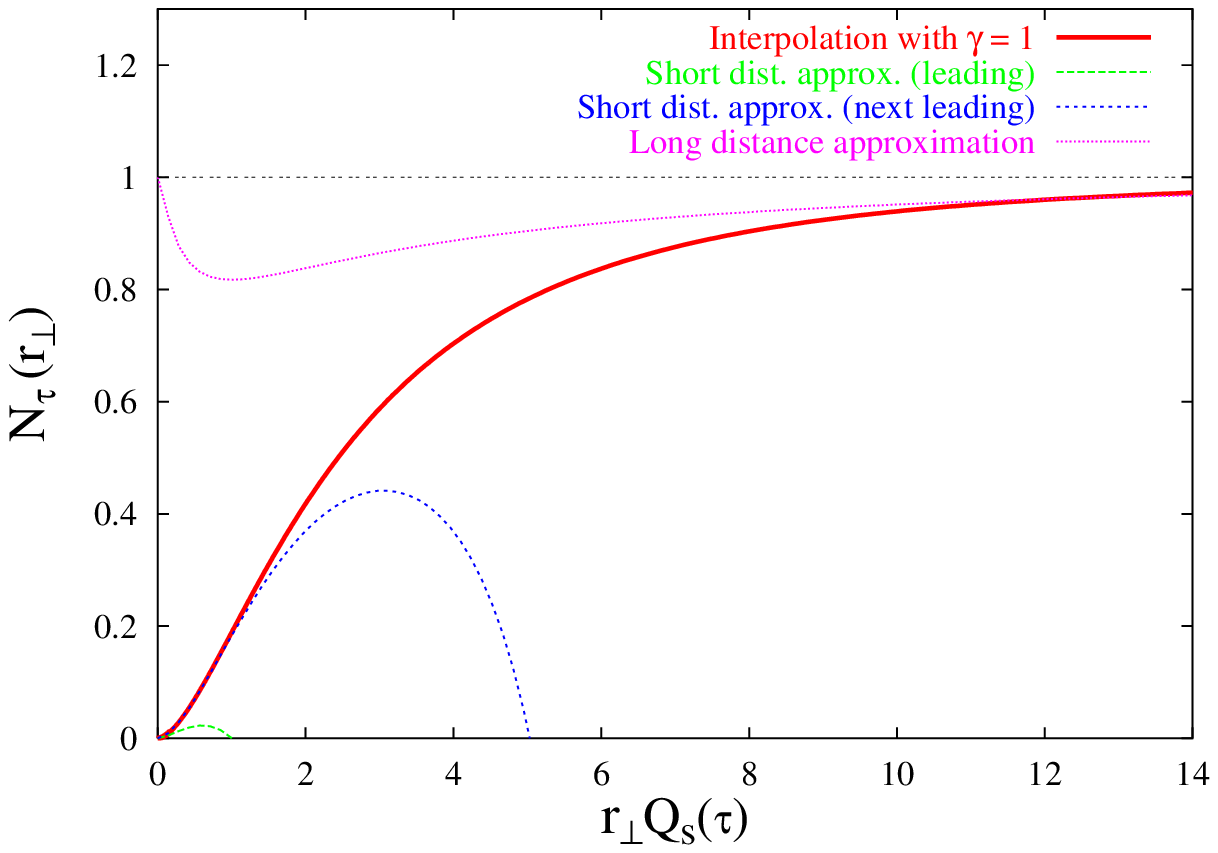}
  }
 \caption[]{%{}\\
      %\hspace*{1.5cm}
      The scattering amplitude $1-S_\tau(r_\perp)$ for $\gamma=1$ 
       plotted as a function of $r_\perp Q_s(\tau)$. Red solid line: 
 numerical evaluation of eq.~(\ref{barS});
green dashed line: leading--order short distance 
      approximation, eq.~(\ref{Stau_short_gamma_1});
 blue dotted line: next--to--leading--order
short distance approximation, eq.~(\ref{short_dist_app});
 magenta dotted line: 
      long distance approximation, eq.~(\ref{Stau_long}).}
\label{F7}
\end{figure}

Let us finally return to the case $\gamma=1$, 
and show that, in this case, the behaviour of $S_\tau(\rr)$ around the
saturation scale can be studied  analytically.
To this aim, it is preferable to return to the expression of $S_\tau(\rr)$
in eq.~(\ref{Stau-MOM}) [with 
$\bar\lambda_{\rm y}(k_{\perp})$ as given by eq.~(\ref{bar2})],
and perform first the momentum integration there, for fixed y.
This generates a  modified Bessel function (see the Appendix for more details),
which is then expanded in an infinite series, and the result is
integrated over y term by term. The final result reads:
\BQA\label{final_bar1}
\hspace*{-.5cm}
%S_\tau(\rr)\bigg |_{\gamma=1}& =&{\rm e}^{-\Omega_\tau(\rr; \gamma=1)}\, ,\\
\Omega_\tau(\rr)\bigg |_{\gamma=1}&=&
\frac{4}{c}\sum_{n=1}^\infty 
\frac{(Q_s^2(\tau)r^2)^n}{(n!)^2\, 4^n \, 2n}
\left\{
\left(\frac{1}{2n}+\psi(n+1)+\ln 2 \right)-\frac{1}{2}\ln (Q_s^2(\tau)r^2)
\right\}.
\EQA
By using this exact series expansion, one can study the approach
towards saturation from shorter distances.
Namely, for $\rr\ll 1/Q_s(\tau)$, 
the dominant contribution comes from the $n=1$ term in eq.~(\ref{final_bar1}), 
which yields 
\BQA
\!\!\!\!
S_\tau^{\rm \, short}(\rr)\bigg |_{\gamma=1}\simeq 
\exp\left\{
-\frac{1}{4c} 
\,Q_s^2(\tau)r^2
\left[ \ln\frac{1}{Q_s^2(\tau)r^2}+\Big(1+2\psi(2)+2\ln 2\Big) 
\right]\right\} .\label{short_dist_app}
\EQA
As compared to the leading--log result in eq.~(\ref{Stau_short_gamma_1}),
the  expression above displays also
the subleading term under the logarithm.
Clearly, when $\rr$ approaches $1/Q_s(\tau)$ from below, 
this formally ``subleading'' term gives the dominant behaviour.

In Figs. \ref{F6} and \ref{F7}, we represent the results of the
numerical evaluation of the  scattering amplitude 
${\cal N}_\tau(\rr)=1-S_\tau(r_\perp)$, eq.~(\ref{barS}), for
two values of the anomalous dimension: $\gamma=0.64$ (Fig.  \ref{F6})
and $\gamma=1$ (Fig.  \ref{F7}). In the same plots, we display, for
comparison, the corresponding predictions of the various approximations
derived previously in this subsection.

\section{Conclusions and perspectives}

In this paper, we have proposed a simple Gaussian approximation for
the weight function for the effective Colour Glass description of
the small--$x$ gluons in the hadron wavefunction. The main virtue
of this approximation is its simplicity. The new weight function 
is as simple to use in practice as that of
the original McLerran--Venugopalan model,
but it improves over the latter by including the correlations among the
colour sources associated with quantum evolution towards small $x$ and
gluon saturation. This ensures, in particular, the removal of the
infrared divergences inherent to the MV model, without invoking
non--perturbative physics: Gauge--invariant observables come out
infrared finite because of the colour neutrality of the saturated gluons
over the (relatively short) scale $1/Q_s(\tau)$.
More generally, this Gaussian weight function 
implements the BFKL dynamics (for both the gluon
distribution, and the dipole--hadron scattering)
at high transverse momenta, and gluon saturation at low 
transverse momenta, with the change between these two regimes
taking place at the saturation scale, as it should.

In its simplest form, the kernel of the Gaussian is a scaling function,
eq.~(\ref{bar2}), which is strictly valid
up to momenta $\kk\simle Q_s^2(\tau)/\Lambda_{QCD}$, and involves
three dimensionless parameters: the anomalous dimension $\gamma$, the
logarithmic derivative of the saturation momentum
$\lambda \equiv d\ln  Q_s^2(\tau)/d\tau$ (throughout this paper, we have rather
denoted this quantity
as $\lambda = c\bar\alpha_s$ ; see, e.g., eq.~(\ref{Qs0})), and
the initial rapidity $\tau_0 = \ln(1/x_0)$ which appears when the
saturation momentum, eq.~(\ref{Qs0}), is rewritten as $Q_s^2(x)
= Q^2_0 {\rm e}^{\lambda (\tau-\tau_0)} \equiv Q^2_0 (x_0/x)^{\lambda}$,
with $Q_0$ some convenient scale of reference (e.g., $Q_0= 1$ GeV).
Of course, within the present approximations, the values of $\gamma$
and $\lambda$ are determined by the leading--order BFKL dynamics. But
since we expect geometric scaling and the exponential rise of the
saturation scale with $\tau$ to be more general than just LO BFKL,
we can also treat these quantities as free parameters, to be fitted 
from the data, or taken from some more refined calculations.
For instance, the fits to the electron--proton deep inelastic
scattering data at HERA using the ``saturation model'' \cite{GBW99,BGBK} provide
a value $\lambda\simeq 0.3$, which, remarkably, appears to be consistent
with a recent calculation \cite{DT02} of the saturation scale using the NLO
BFKL dynamics \cite{NLBFKL} (with  the RG--improved kernel proposed
in Refs. \cite{Salam99}).

The Gaussian weight function that we have constructed and justified
 here can be immediately
used to supplant the MV model in its applications to the physics at 
small $x$.

For the problems involving only one high--density hadronic
system (one large 
nucleus\footnote{For applications involving large nuclei, one needs
also to compute, or parametrize, the $A$--dependence of the saturation
scale. One expects a power--law increase, $Q_s^2(\tau,A)
\propto A^\delta$, with $\delta\simeq 1/3$ (see, e.g., 
\cite{eAscaling,KLM02}).}, or a very energetic hadron), the calculations
can be performed quasi--analytically, since the solution to the
corresponding classical field equations is known in analytic form
(cf. Section 2.1). This includes various studies of 
proton--nucleus ($pA$) collisions \cite{KM98,DM01,DJ01,GJ01},
or of ultra--peripheral nucleus--nucleus ($AA$) collisions \cite{GP01}. 
Also, the expression for the dipole--hadron scattering amplitude
that we have obtained in this context, eq.~(\ref{barS}), can be used
in analyses of the DIS data at HERA,
as an alternative to the more phenomenological parametrizations proposed
in Refs. \cite{GBW99,BGBK}. It would be interesting, in this respect,
to compare eq.~(\ref{barS}) with the numerical solution of the BK
equation \cite{LT99,AB01,Motyka,LL01}, or with the results of the
lattice simulations of  the full RGE \cite{HWR} (the latter should be
soon available \cite{HWpers}).
We expect our analytic formula (\ref{barS}) to be numerically
close to such more complete calculations, 
while being much simpler to use in practice,
and allowing for the freedom to choose (or fit) the
values of the parameters $\gamma$ and $\lambda$
(in the numerical calculations alluded to above, these parameters
are rather fixed to their respective BFKL values, by the dynamics of the
relevant equations).

For applications of the CGC effective theory to heavy ion collisions, 
the classical field equations can be numerically solved on a lattice,
as in Refs. \cite{KV}, and their results averaged over the colour
charge distributions in the incoming nuclei, with the Gaussian
weight function. As compared to the MV model, the use of the new Gaussian 
in such applications would eliminate the
sensitivity to the poorly known non--perturbative physics.
Still within the context of $AA$ collisions, we note that the
 simple analytic form we have derived for 
the unintegrated gluon distribution, eq.~(\ref{barphi}), can be directly
implemented in analytic calculations of the multiparticle production, like
those in Refs. \cite{KLM02,KNL}.

\vspace*{.5cm}
\section*{Acknowledgments}

We would like to thank Miklos Gyulassy for having triggered this work
with his incentive remarks, and Al Mueller for illuminating discussions
on the implications of color neutrality for the calculation of 
the gluon distribution function. We are grateful to Fran\c cois Gelis
for helping us with the numerical calculations, and for a very careful
reading of the manuscript, with many useful remarks.

This manuscript has been authorized under Contracts No. DE-AC02-8-CH10886
and No. DE-AC02-76CH0300 with the US Department of Energy.

\appendix

\section{Appendix} %: Derivation of eq.~(\ref{final_bar1})}
\setcounter{equation}{0}

In this Appendix, we derive the analytic expression (\ref{final_bar1})
for the exponent of the $S$-matrix element, which refers to 
eq.~(\ref{barS}) with $\gamma=1$. 
We first perform the momentum integration in eq.~(\ref{Stau-MOM}),
where $\lambda_{\rm y}(\kk)$ is now replaced by its interpolation
in eq.~(\ref{bar2}) (with $\gamma=1$). 
 It is useful to decompose this latter as follows:
\be
\frac{1}{\kk^4}\, \bar\lambda_{\rm y}(\kk)\Bigg |_{\gamma=1} %= 
%\frac{1}{\pi}\cdot \frac{Q_s^2({\rm y})}{\kk^2(\kk^2+Q_s^2({\rm y}))}
=\frac{1}{\pi}\left[\frac{1}{\kk^2}-\frac{1}{\kk^2+ Q_s^2({\rm y})}\right].
\ee
When this function
is inserted in the integrand of eq.~(\ref{Stau-MOM}), its product with 
 $1-{\rm e}^{ik_\perp\cdot r_\perp}$ is naturally
decomposed into four terms. Since the overall integral is finite, 
we are allowed to evaluate these four 
terms separately, provided we use the same regularization scheme. 
All these terms are represented by a simple  
integral (after angular integration, $r=|\rr|$):
\BQ\label{finite}
\int_0^\infty dk\, J_0(kr)\, \frac{k}{k^2 + P^2} \, = \, K_0(Pr)\, ,
\EQ
where $K_0(z)$ is the modified Bessel function (with
$\psi(z) = d\ln \Gamma(z)/dz$)
\BQ\label{mod_Bessel}
K_0(z) =  -\ln z +(\ln 2 -\gamma)+
\sum_{n=1}^\infty \frac{1}{(n!)^2}\left(\frac{z}{2}\right)^{2n}
\left(\psi(n+1)-\ln \frac{z}{2}\right).
\EQ
For example, the term $\int (d^2\kk/\kk^2)$, which has both ultraviolet and 
infrared divergences, can be regularized as
$\int 2\pi kdk\, J_0(k / \Lambda)/(k^2 + \mu^2)$, 
with $\Lambda$ and $\mu$ being
the UV and IR cutoffs, respectively,
and thus brought into the form of eq.~(\ref{finite}).
When adding the contributions of the four terms,  all divergences
cancel out, as they should, and, in the limit $\Lambda\to \infty$ and
$ \mu\to 0$, we are left with the following result:
\BQA\label{exp_after_mom}
\Omega_\tau(\rr)\Big\vert_{\gamma=1}=2\bar\alpha_s \int^\tau_{-\infty}d {\rm y}\, 
\Big\{\ln(Q_s({\rm y})r) - (\ln 2 -\gamma) + K_0(Q_s({\rm y})r)\Big\}.
\EQA
The integral over the rapidity y can be now
 performed, by using the energy dependence 
of $Q_s({\rm y})$ given in eq.~(\ref{Qs0}) :
\BQA
\Omega_\tau(\rr)\Big\vert_{\gamma=1}& =& \frac{4}{c}
\int_0^{Q_s(\tau)r}\frac{d\zeta}{\zeta}\Big\{\ln \zeta - (\ln 2 -\gamma) 
+ K_0(\zeta)\Big\}\qquad [\ \zeta=Q_s({\rm y})r\ ]\NN
&=& \frac{4}{c}
\int_{0}^{Q_s(\tau)r}d\zeta
\sum_{n=1}^\infty \frac{1}{(n!)^2}\frac{1}{\zeta}\left(\frac{\zeta}{2}\right)^{2n}
\left\{\psi(n+1)-\ln \frac{\zeta}{2}\right\}.
\label{full}
\EQA
In the last line, we have used eq.~(\ref{mod_Bessel}).
By integrating this series 
term by term, one obtains the final result in eq.~(\ref{final_bar1}).

\end{document}